\documentclass[fleqn,usenatbib]{mnras}

\usepackage{newtxtext,newtxmath}

\usepackage[T1]{fontenc}
\usepackage{siunitx}
\DeclareRobustCommand{\VAN}[3]{#2}
\let\VANthebibliography\thebibliography
\def\thebibliography{\DeclareRobustCommand{\VAN}[3]{##3}\VANthebibliography}

\usepackage{graphicx}	
\usepackage{amsmath}	
\usepackage[inline]{enumitem}

\title[]{Evolving Solar Wind Flow Properties of Magnetic Inversions Observed by \textit{Helios}}

\author[A. R. Macneil et al.]{
Allan R. Macneil$^{1}$\thanks{E-mail: a.r.macneil@reading.ac.uk (ARM)},
Mathew J.~Owens$^{1}$,
Robert T.~Wicks$^{2}$,
Mike~Lockwood$^{1}$
\\
$^{1}$Department of Meteorology, University of Reading, Reading, UK\\
$^{2}$Department of Mathematics, Physics and Electrical Engineering, Northumbria University, Newcastle-upon-Tyne, UK
}

\date{Accepted XXX. Received YYY; in original form ZZZ}

\pubyear{2020}

\begin{document}
\label{firstpage}
\pagerange{\pageref{firstpage}--\pageref{lastpage}}
\maketitle

\begin{abstract}

In its first encounter at solar distances as close as $r=\SI{0.16}{AU}$, \textit{Parker Solar Probe} (PSP) observed numerous local reversals, or inversions, in the heliospheric magnetic field (HMF), which were  accompanied by large spikes in solar wind speed.
Both solar and \textit{in situ} mechanisms have been suggested to explain the existence of HMF inversions in general. 
Previous work using \textit{Helios} 1, covering 0.3--\SI{1}{AU}, observed inverted HMF to become more common with increasing $r$, suggesting that some heliospheric driving process creates or amplifies inversions. 
This study expands upon these findings, by analysing inversion-associated changes in plasma properties  for the same large data set, facilitated by observations of `strahl' electrons to identify the unperturbed magnetic polarity.
We find that many inversions exhibit anti-correlated field and velocity perturbations, and are thus characteristically Alfv\'enic, but many also depart strongly from this relationship over an apparent continuum of properties. Inversions depart further from the `ideal' Alfv\'enic case with increasing $r$, as more energy is partitioned in the field, rather than the plasma, component of the perturbation. This departure is greatest for inversions  with larger density and magnetic field strength changes, and characteristic slow solar wind properties. 
We find no evidence that inversions which stray further from `ideal' Alfv\'enicity have different generation processes from those which are more Alfv\'enic.
Instead, different inversion properties could be imprinted based on transport or formation within different solar wind streams.

\end{abstract}




\section{Introduction}\label{sec:intro}
`Inversions' in the polarity of the  heliospheric magnetic field (HMF) occur when a magnetic flux tube has observed polarity opposite to that where it connects back to the Sun. Inversions can occur when the field becomes locally reversed, and have been observed across a range of heliospheric distances and latitudes, with durations ranging from minutes to longer than a day  \citep[e.g.,][]{Kahler1994,Balogh1999,Crooker2004,Yamauchi2004,Matteini2014,Owens2020}. 
Parker Solar Probe \citep[PSP,][]{Fox2016} observations of inversions close to the Sun have renewed interest in their origin and evolution. \textit{PSP} found magnetic inversions to be ubiquitous in the inner (down to $\SI{0.16}{AU}$) heliosphere. These inversion durations range from some seconds to hours \citep{TDW2020}.
The change in the field orientation is matched with an oppositely-directed spike in the bulk velocity \citep[][]{Bale2019,Kasper2019}. The anti-correlated field and velocity change, as well as the relatively stable density and magnetic field strength, during these inversions suggests that they are Alfv\'enic, and the anti-sunward spike in velocity indicates outward propagation. These Alfv\'enic magnetic inversions are often referred to as `switchbacks' (however this term is also at times used more broadly as a synonym of magnetic inversion). Alfv\'enicity has also been noted as a key characteristic of inversions observed in the fast solar wind elsewhere in the heliosphere, e.g., at \SI{0.3}{AU} \citep{Horbury2018} and \SI{>1}{AU} \citep{Matteini2014}.

The origins of magnetic inversions are not well understood and may provide insight into the processes which generate the solar wind. 
Following the opening of a coronal magnetic loop through interchange reconnection \citep[a key process in many solar wind models][]{Fisk2003,Antiochos2011}, the newly-opened magnetic flux tube will have an S-shaped kink embedded within it \citep[see e.g., Figure 1 of][]{Crooker2002}.
This  kinked magnetic structure propagates upwards through the corona at approximately the local Alfv\'en speed. Different modelling approaches have demonstrated that these kinks may survive to reach solar distances observable by PSP \citep{Zank2020} or further \citep{Tenerani2020} as magnetic inversions.  
This model for inversion formation has been applied in the past to use inversions as a solar wind tracer of coronal reconnection \citep[e.g.,][]{Crooker2002,Crooker2004,Owens2013b}. Recently, the ubiquity of  PSP switchbacks has been cited as evidence supportive of the interchange model of the solar wind \citep[][]{Fisk2020}.

Rather than being initially created at the Sun, inversions in the HMF could form as a result of numerous \textit{in situ} processes during solar wind transport between the Sun and the observer. Modelling by \cite{Squire2020} has demonstrated that switchbacks could gradually develop in the solar wind as a result of the growth of Alfv\'enic turbulent fluctuations due to solar wind expansion. 
Earlier work by \cite{Landi2005,Landi2006} showed that velocity shears can sufficiently distort the field to form inversions through interaction with  outward-propagating large scale Alfv\'en waves (i.e., low frequency Alfv\'enic turbulence) which create a misalignment between the radial solar wind flow and the magnetic field direction. 

The large scale structure of the solar wind has also been explored as a mechanism for \textit{in situ} inversion generation.
Similar to the model of \cite{Landi2006}, \cite{Owens2018,Owens2020} describe how magnetic flux tubes threaded by solar wind speed shears can form inversions at the location of the shear, due to the flux tube convecting radially  at different speeds along its length. This process can also be related to solar wind formation processes, as such shears along a flux tube can arise as a result of interchange reconnection at the tube's footpoint, leading to a time variation in the plasma source at the flux-tube base. 
\cite{Owens2020} found evidence for such a formation process from analysis of large scale (duration $>\SI{0.3}{days}$) inversions at \SI{1}{AU}, which were found to statistically coincide with both velocity shears, and changes in heavy ion composition - a signature of connectivity to different source regions.
Inversions could also be driven through the draping of the HMF over ejecta associated with solar wind transients such as interplanetary coronal mass ejections (ICMEs) or plasma blobs \citep{Lockwood2019}.
The above driving of inversions through distortion by streams and structures cannot be expected to produce inversions for which a typical Alfv\'enic field and flow correlation exists. Stream shear will produce a structured velocity field at the inversion, while draping will involve compression.

\cite{Macneil2020inv} performed a statistical study of inverted HMF occurrence using data from the \textit{Helios} 1 spacecraft. They used the direction of the suprathermal beam of electrons known as the `strahl', which predominantly travels away from the Sun along the HMF \citep{Feldman1975,Hammond1996,Pierrard2001} to discriminate between locally inverted HMF and true sector reversals.
The primary result of this study was that the occurrence of inverted HMF increases between 0.3 and \SI{1}{AU}. If inversions generated at the Sun tend to naturally decay with distance, then this result supports the view that a significant fraction of inversions are actively driven in the heliosphere, possibly by one or more of the  processes discussed above. Velocity shears \citep[][]{Landi2006} or other \textit{in situ} processes could also be acting to amplify inversions which are generated at the Sun, contributing to this trend. As a corollary, since this trend is only reported between 0.3 and \SI{1}{AU}, it is thus possible that inversions observed within \SI{0.3}{AU} by PSP may have different causes to those found at 0.3--\SI{1}{AU}.

A further result of \cite{Macneil2020inv} was that inversions did not exhibit a particular bias in the direction of deflection of the azimuthal magnetic field.  For this reason,  a driving process such as turbulence was favoured as an explanation for the increase in inverted HMF. However, subject to a range of caveats, other processes could still reasonably contribute.
Recent studies on the near-Sun switchbacks observed by PSP indicate that they occur in clusters \citep{TDW2020}, within which the deflections tend to be in the same direction \citep{Horbury2020}, suggesting a coronal origin is more likely for those inversions.

PSP-observed switchbacks are mostly Alfv\'enic disturbances, exhibiting correlated velocity and magnetic field variation, which propagate along the field in the anti-sunward direction. For an Alfv\'enic fluctuation: 
\begin{equation}\label{eq:dv}
    \delta \mathbf{v} = \pm\alpha \delta\mathbf{B}
\end{equation}
where $\alpha = \sqrt{r_A} v_A/B_0$, $r_A$ is the Alfv\'en ratio (the ratio of kinetic to magnetic energy in the fluctuations), $v_A$ is the local Alfv\'en speed, and $B_0$ is the magnitude of the unperturbed magnetic field \citep[e.g.,][]{Matteini2014,Horbury2018}. 
In the solar wind, $r_A$ is typically $<1$. $r_A$  decreases with solar distance in the inner heliosphere, reaching around 0.4--0.5 for fast wind and around 0.25 for slow wind near \SI{1}{AU}, with a degree of spread \citep{Marsch1990,Bruno2005,Borovsky2012}.
The `$+$' sign in Equation \ref{eq:dv} applies for inward HMF sectors, and the `$-$' sign for outward sectors, for fluctuations that propagate in the anti-sunward direction. In Section \ref{sub:invmeth}, we rectify the observations such that the outward sector case applies at all times, and so we continue using only the `$-$' form of this expression.
Assuming that $|B| = B_0$ throughout the fluctuation (i.e., the field vector is rotating on a spherical surface), then $\delta \mathbf{v}$ can be expressed as a function of the deflection angle of the field. If we further assume this rotation to lie in the $R$--$T$ plane (in RTN coordinates) then the field rotation will take the form of an azimuthal deflection; $\Delta \phi$.
For the purpose of this study, $|\Delta \phi| > \SI{90}{\degree}$  indicates inverted HMF, as this is the threshold beyond which the field is directed opposite to its unperturbed direction.
Assuming that the unperturbed field orientation is the nominal Parker spiral direction, then 
$\Delta \phi = \Delta \phi_P \equiv \phi - \phi_{\mathrm{P}}$, where $\phi$ is the azimuthal field angle, and $\phi_P$ is the Parker spiral angle.
The change in the velocity component parallel to the nominal Parker spiral direction, $v_{\parallel}$, is then 
\begin{equation}\label{eq:dvp}
    \delta v_{\parallel} = \sqrt{r_A}v_A[1-\cos(\Delta \phi_P)],
\end{equation}
and the component orthogonal to the Parker spiral direction in the $R$-$T$ plane is 
\begin{equation}\label{eq:dvop}
    \delta v_{\perp} = -\sqrt{r_A} v_A\sin(\Delta \phi_P).
\end{equation}
Equation \ref{eq:dvp} can be understood intuitively by noting that when $\mathbf{B}$ initially follows the Parker spiral direction, any $|\Delta \phi_P| >0$ is a decrease in the Parker spiral component of $\mathbf{B}$ (as long as its magnitude is constant). Given Equation \ref{eq:dv}, this must then produce an increase in the corresponding velocity component, $v_{\parallel}$. In cases where the ideal Parker spiral does not adequately describe the background magnetic field (i.e., there is an angular offset between the true background field angle $\phi_0$ and $\phi_P$) the estimated values of both $\Delta \phi_P$ and $\delta v_{\parallel (\perp)}$ will be altered. The impact of this on attempts to characterise Alfv\'enic inversions are assessed in Appendix \ref{app:angles}.

Alfv\'enicity is a key observational feature of inversions, particularly for the near-Sun switchbacks observed by PSP. More generally, the relationship between field and velocity perturbation during inversions is useful to investigate in detail, given that some suggested mechanisms for generating inversions above will not necessarily produce an Alfv\'enic relation. 
Alfv\'enic inversions were shown to exist in fast solar wind in the inner portion of the \textit{Helios} orbit (around \SI{0.3}{AU}) by \cite{Horbury2018}, and in high-latitude fast wind at distances \SI{>1}{AU} by \cite{Matteini2014}, but further insight on inversion properties over a continuous range of distances and in other solar wind conditions can still be extracted from the \textit{Helios} data set.
In this paper we extend this work and the work of \cite{Macneil2020inv}; performing  analysis of changes in proton velocity, density, and magnetic field magnitude  as a function of magnetic field azimuthal deflection. We do so over the full time range of the \textit{Helios} 1 data set, enabled by the magnetic sector information provided by the strahl. 
This allows us to analyse how the Alfv\'enic, or otherwise, nature of magnetic inversions changes with distance from the Sun, and for different solar wind conditions.
Section \ref{sec:datmeth} describes the \textit{Helios} 1 data used in this study, and the methodology for identifying inverted HMF and calculating the deflection angle. 
Section \ref{sec:res} describes the results, which are discussed in the context of the production of HMF inversions in Section \ref{sec:disc}. We draw conclusions in Section \ref{sec:conc}. Two appendices are included which report on additional details of the methodology.

\section{Data and Methodology}\label{sec:datmeth} 
\subsection{Helios Data}

We use data from \textit{Helios} 1, covering a distance range of 0.3--\SI{1}{AU}, over the years 1974--1981. The data and methodology sections of \cite{Macneil2020sunward,Macneil2020inv} describe this data in detail, and we summarise here. 
Electron and proton data for the study are obtained on \SI{40}{s} cadence from the \textit{Helios} Plasma Experiment (E1). 
Proton measurements were made with the E1-I1 ion instrument. We use fitted moments of the proton velocity distribution function (VDF) from the reanalysed \textit{Helios} `corefit' data set \citep{Stansby2018}. 
Electron VDF measurements were made with the E1-I2 electron instrument. 
Vector magnetic field data from the \textit{Helios} Magnetic Field Experiment (E2) are time-averaged to the same \SI{40}{s} cadence, and combined with the electron VDFs to produce electron pitch angle distributions (PADs) from which the strahl direction is extracted  to identify the HMF morphology \citep[details of the processing to produce the PADs can be found in][]{Macneil2020sunward}.

Removal of a portion of the data from our study is necessary because the limited field of view of E1-I2 regularly causes the main component of the strahl electron beam to fall outside of the detector aperture. We follow the same removal of data described in \cite{Macneil2020inv} \citep[adapted from methods in][]{Maksimovic2005,Stverak2009}; discarding samples where the out-of-ecliptic magnetic field component, $B_z$, is large relative to the total field magnitude $B$: $B_z/B > 0.156$. 
We also exclude a further portion of data where properties of the electron PADs do not allow a strahl direction to be determined \citep[details in][]{Macneil2020sunward}.

\subsection{Inversion Identification and HMF Deflection Angle}\label{sub:invmeth}
In typical solar wind conditions, strahl electrons travel anti-sunward along the field. Intervals where the strahl instead travels sunward are thus identifiable as local HMF inversions. 
We determine the strahl orientation for each electron PAD using the method detailed in \cite{Macneil2020sunward}, which also identifies instances of bidirectional strahl and strahl drop outs which we then remove from this study. 
Since the unperturbed HMF follows the Parker spiral, uninverted HMF which is positive (negative) relative to the expected Parker spiral direction will have a  parallel (anti-parallel) oriented strahl beam. Inverted HMF which is positive (negative) along the expected Parker spiral direction will have anti-parallel (parallel) strahl. 
This information, summarised in Figure 1 of \cite{Macneil2020inv}, is used to tentatively classify all HMF samples which have a detectable mono-directional strahl as either uninverted or inverted.

To mitigate the effect of possibly mis-identified inverted HMF on our results, 
we apply a further restriction on the samples which are categorised as inverted based on the above strahl procedure. 
This is motivated by the existence of a subset of inverted HMF samples which show evidence of mis-identification, detailed in Appendix \ref{app:correction}. 
To remove these, we select for inverted HMF which takes the form of a clear deviation from the background magnetic field direction, i.e., a temporary reversal in the field. We do so by computing the magnetic field component along the nominal Parker spiral direction, $B_P$, for all samples in a 12 \SI{}{hr} window centred on each tentative inverted HMF sample. We use a \SI{12}{hr} window to match that used to characterise the background solar wind in the majority of the results below. For each window, we compute the modal value of $B_P$; $M_{BP}$.  Tentative inverted samples are only considered to be local inversions, and thus preserved in this study, if they have  the opposite polarity to  $M_{BP}$. A similar method of inversion identification is used in place of strahl measurements by e.g., \cite{Badman2020}.
Tentative inverted samples where $B_P$ has the same polarity as $M_{BP}$ are discarded from the study. 
This restriction effectively removes the set of inverted HMF samples which are likely to be falsely-identified inversions.
A consequence of this procedure is that inverted HMF which does not take the form of a deviation from a clearly-defined background magnetic field (such as large scale inversions, inversions which occur near the heliospheric current sheet, or inversions which are indicated by only strahl and not a magnetic field deflection) are also removed.
It is difficult to disentangle these real inversions from the falsely identified ones, so we take the cautious approach of removing them all. Following these steps, a total of 5620 valid inverted HMF samples remain, out of 197972 valid samples overall.
Results and  conclusions of this study thus specifically apply to inverted HMF which takes the form of a temporary reversal relative to the background magnetic field.
Consequences of this restriction are discussed further in Section \ref{sub:caveats} and Appendix \ref{app:correction}.

Having classified samples into inverted and uninverted HMF, we  produce estimates of  $\Delta \phi_P$ from Section \ref{sec:intro} by subtracting the HMF azimuthal angle, $\phi$, from the nominal Parker spiral direction. We rectify $\Delta \phi_P$ by \SI{180}{\degree} for any samples with anti-parallel strahl, since uninverted HMF with this strahl alignment will belong to the negative magnetic sector. Thus, assuming a perfect Parker spiral, all unperturbed field will have $\Delta \phi_P=\SI{0}{\degree}$, and all inverted field will have $|\Delta \phi_P|>\SI{90}{\degree}$. This approach allows us to study relationships between plasma properties and deflection angle, even for inversions, for which the deflection angle would be ambiguous at  $|\Delta \phi_P|>\SI{90}{\degree}$ without rectification. This step is what enables the analysis of the entire \textit{Helios} 1 data set, without splitting into individual streams or sectors.

\section{Results}\label{sec:res}

\begin{figure*}
    \centering
    
    \includegraphics[clip,trim={0 6cm 0 0},width =0.35\textwidth]{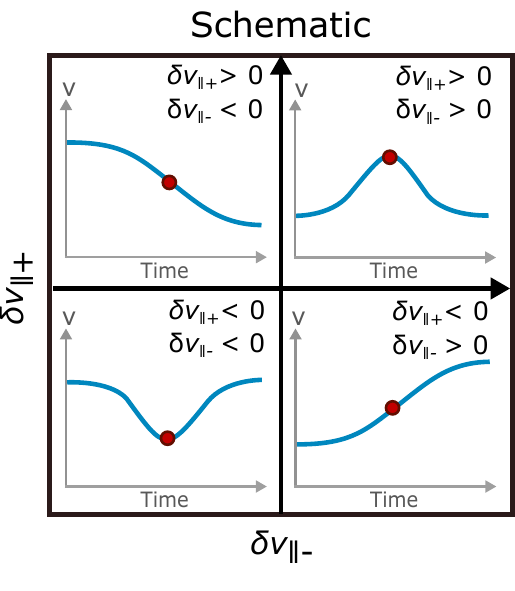}
    \includegraphics[clip,trim={0 0cm 3cm 0},width=0.53\textwidth]{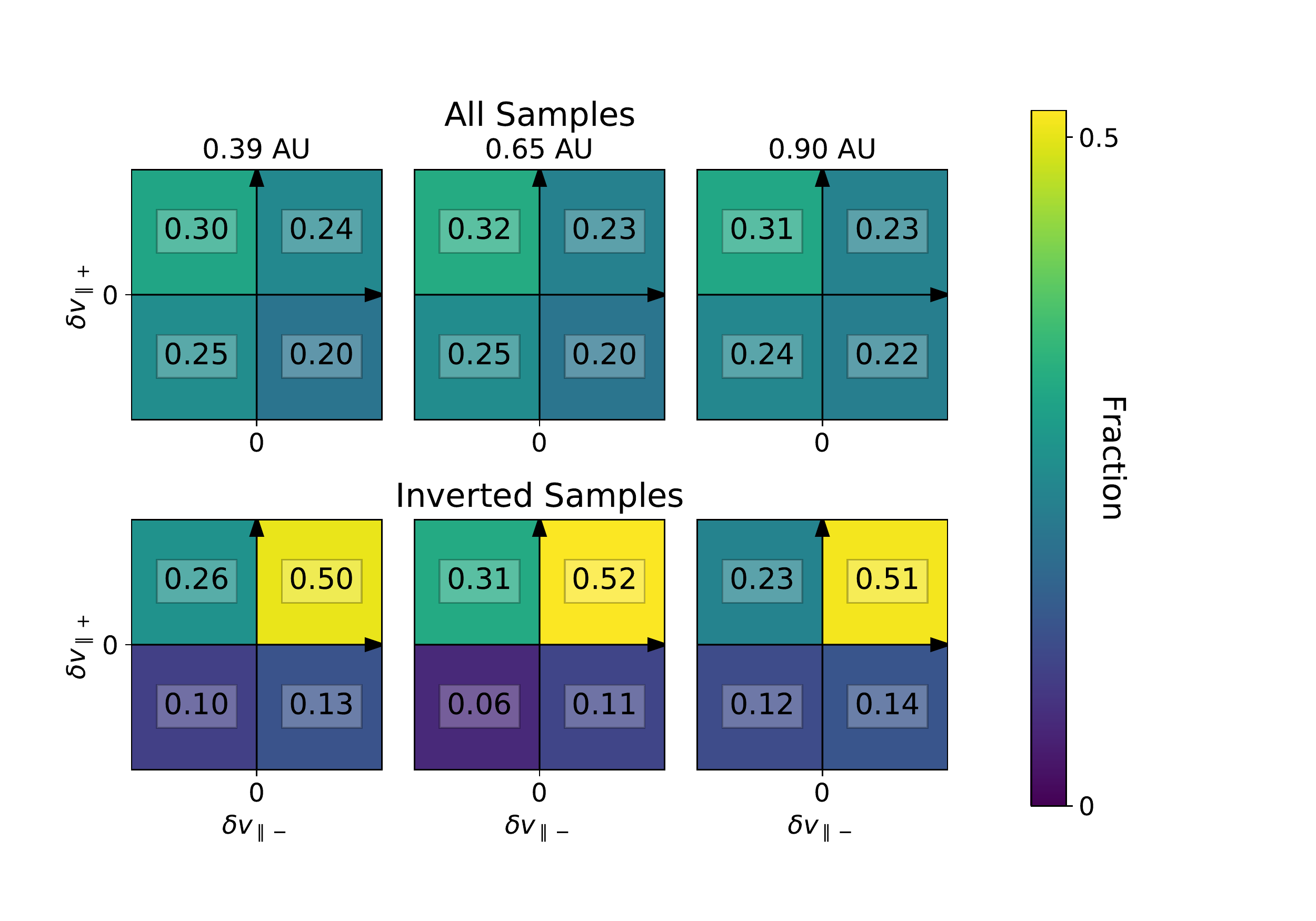}
    \caption{2-dimensional histogram, or contingency table, of $\delta v_{\parallel+}$ against $\delta v_{\parallel-}$, where $\delta v_{\parallel\pm}(t) $ is calculated from Equation \ref{eq:dpm}. Means are calculated over a \SI{1}{hr} window centred on time $t$. 
    Samples where $|\delta v_{\parallel\pm}|/v_A < \SI{e-3}{}$ are not included in the histograms.
    Each table has 4 bins, corresponding to the 4 quadrants of $\delta v_{\parallel+}$ and $\delta v_{\parallel-}$.
    The schematic to the left illustrates the velocity profiles which are implied by each quadrant, where the red dot represents the time of the sample.}
    \label{fig:dvpm}
\end{figure*}

As described above, HMF inversions are often accompanied by a spike in solar wind velocity, which for an Alfv\'enic perturbation will occur in the velocity component parallel to the unperturbed magnetic field, $v_{\parallel}$. We compute $v_{\parallel}$ as the component of the proton velocity vector projected on to the nominal Parker spiral direction. 
We compute differences between a \SI{1}{hr} rolling averaged $v_{\parallel}$, centred on a time $t$, $\langle v_{\parallel} \rangle_{1\mathrm{\,hr}}(t)$, and the result of the same averaging process for $v_{\parallel}$ centred on the hour before and after:
\begin{equation}\label{eq:dpm}
    \delta v_{\parallel\pm}(t) = \langle v_{\parallel}\rangle_{1\mathrm{\,hr}}(t) - \langle v_{\parallel}\rangle_{1\mathrm{\,hr}}(t \pm \SI{1}{hr}).
\end{equation}
Comparing consecutive hourly averages will smear out enhancements which have duration less than tens of minutes, or greater than 2 hours. 
We use an hourly average as a compromise between capturing smaller scale or shorter duration changes in velocity and reducing the noise in the result which comes from a shorter average. 
Data are classified based upon the sign of $\delta v_{\parallel+}(t)$ and $\delta v_{\parallel-}(t)$. The occurrence is shown as 2-dimensional histograms in Figure \ref{fig:dvpm}. We display histograms for all samples, and for inverted HMF separately, for three radial distance bins centred on $r = 0.39$, 0.65, and \SI{0.90}{AU}. The histograms have one velocity bin in each $\delta v_{\parallel+}$--$\delta v_{\parallel-}$ quadrant, such that each is analogous to a contingency table. A schematic to the left of the figure illustrates the time series profile of $v_{\parallel}$ which is indicated by each quadrant of the table. These are (clockwise from top left) declining, temporary spike, increasing, and temporary dip.
We have excluded data for which $|\delta v_{\parallel\pm}/v_A|\leq\SI{e-3}{}$, where $v_A$ is the local Alfv\'en speed, such that small changes in velocity are considered to be zero, and not included. This procedure removes between 0.5--\SI{1.5}{\percent} of the relevant data points for the inverted HMF histograms, but makes little difference as to the relative proportions in each quadrant.

The $\delta v_{\parallel\pm}$ contingency tables for all valid samples are shown in the top row of Figure \ref{fig:dvpm}. The declining velocity quadrant is the most common at all distances, which is consistent with the overall declining velocity profile of the solar wind due to rarefactions \citep[see e.g.,][]{Gosling1999}. The least common quadrant corresponds to increasing velocity, which we expect for the same reason. 
Contrast these results with the $\delta v_{\parallel\pm}$ contingency tables for locally inverted HMF. At all distances, the velocity spike quadrant corresponds to over half of the samples. The increase in the velocity spike quadrant in comparison to the uninverted case comes at the expense of all velocity quadrants.
The declining quadrant in the middle distance bin for inverted samples is noticeably greater than that for the other distance bins. This irregularity may be a result of some sampling effect, since the middle bin has the fewest samples. Through this section, the middle distance bin will often show some degree of discrepancy relative to the other two, more well-populated, bins.

The clear velocity spike signal for inversions is quite remarkable given the point noted above regarding the inversion duration. The prevalence of these spikes seen with the $\pm \SI{1}{hr}$ difference method indicates that a number of these inversions have velocity spikes which are resolvable on hourly scales. 
Estimating $\delta v_{\pm}$ using larger rolling average windows and time offsets (up to 12 hours) in Equation \ref{eq:dpm} produces results consistent with those above. While the declining quadrants in each histogram become more prominent, an enhancement in the velocity spike sector is present for  the inverted HMF histograms, in the two innermost distance bins, over their uninverted counterparts. The histogram for the outer distance bin is the first to lose the clear spike signature as the averaging and offset times are increased.

\begin{figure*}
    \centering
    \includegraphics[clip,trim={0 10.05cm 0 0},width=0.8\textwidth]{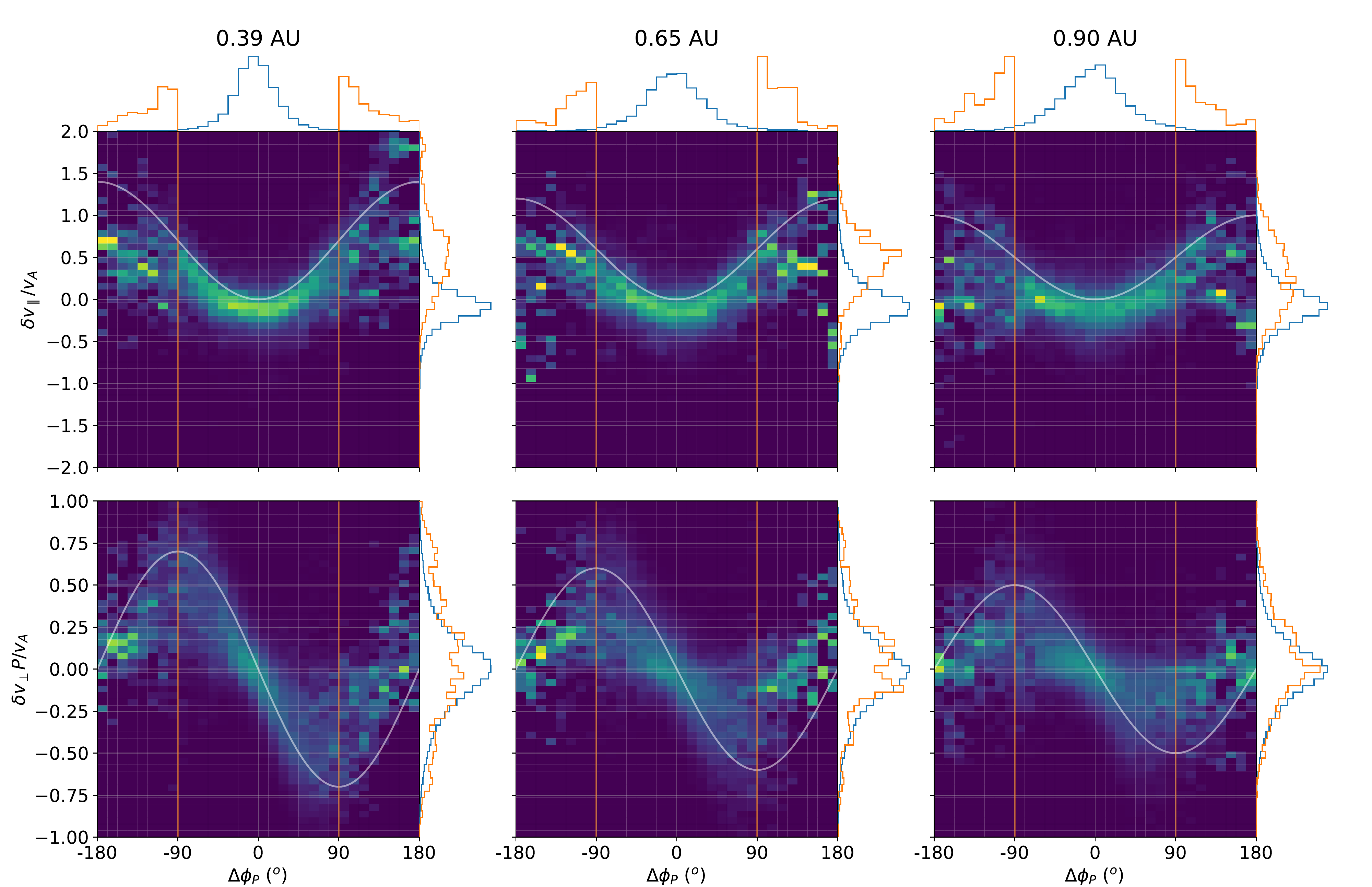}
    \caption{2-dimensional histograms of $\delta v_{\parallel}/v_A$ (top row) and $\delta v_{\perp}/v_A$ (bottom row) against $\Delta \phi_P$  for three radial distance bins centred on $r = [0.39, 0.65, 0.90] \,\SI{}{AU}$ respectively. $\delta v_{\parallel}$ and $\delta v_{\perp}$ are calculated using the form of Equation \ref{eq:dX} with a time average $\mathrm{Y}=\SI{12}{hr}$.
    Each histogram is normalised such that the sum of values within each column $=1$. Orange lines at $\Delta \phi_P =\pm\SI{90}{\degree}$ denote the boundary between uninverted and inverted HMF. White lines show the theoretical relationship for each parameter as described in Section \ref{sec:intro} (Equations \ref{eq:dvp} and \ref{eq:dvop}) with $\sqrt{r_A} = [0.7,0.6,0.5]$ for the three distance bins respectively.
    Above the top row of panels, the blue histogram shows the distribution of $\Delta \phi_P$ for all samples in the relevant distance bin, while the orange histogram shows the distribution  of $\Delta \phi_P$ for inverted HMF samples only. The orange and blue histograms are normalised independently.  
    To the right of each panel, normalised histograms show the distribution of that panel's $y$-axis parameter for  uninverted and inverted HMF samples combined (blue line) and for inverted samples only (orange line). }
    \label{fig:v_heatmap}
\end{figure*}

Figure \ref{fig:dvpm} shows that inversions preferentially feature a spike in velocity along the Parker spiral direction. However, it does not give insight into the magnitude of the spike, or how it relates to the degree of deflection of the field.
To address this, we define the values $\delta v_{\parallel}$ and $\delta v_{\perp}$ by subtracting rolling averages of $v_{\parallel}$ and $v_{\perp}$, calculated within a given time window, from the \SI{40}{s} data. 
$v_{\perp}$ is computed as the velocity component orthogonal to the nominal Parker spiral direction in the $R$-$T$ plane.
We use the following general formula to compute $\delta v_{\parallel}$ and $\delta v_{\perp}$:
\begin{equation}
    \delta X = X_{\SI{40}{s}} -  \langle X\rangle_{\mathrm{Y\,hr}}
\end{equation}
for parameter $X$, where $Y$ is the duration of the averaging window in hours.
The averaging window is centred on the time stamp of the \SI{40}{s} observation. A \SI{12}{hr} average is used here, roughly corresponding to the expected large scale solar wind structure duration. Results are qualitatively similar for other values of $Y$. 
$\delta v_{\parallel}$ and $\delta v_{\perp}$ are then normalised by the mean Alfv\'en speed over the same time window:
\begin{equation}\label{eq:dX}
    \langle v_A\rangle_{\mathrm{Y\,hr}} = \left\langle \frac{B}{\sqrt{\mu_0 m_p n}}\right\rangle_{\mathrm{Y\,hr}}
\end{equation}
The use of \SI{40}{s} data to estimate $\delta v_{\parallel(\perp)}$ implicitly limits the scope of results acquired with this method to inversions with duration around this time scale or greater. Conversely, the \SI{12}{hr} averaging time for the background velocity is effective for inversions of duration  \SI{<6}{hr}. 

Figure \ref{fig:v_heatmap} shows 2-dimensional histograms of $\delta v_{\parallel}/v_A$ and $\delta v_{\perp}/v_A$  against $\Delta \phi_P$, for the same contiguous radial distance bins as Figure \ref{fig:dvpm}. 
At the top of Figure \ref{fig:v_heatmap}, 1-dimensional histograms of $\Delta \phi_P$ for each distance bin (in blue) show that the data are heavily concentrated around $\Delta \phi_P =0$. For this reason the histograms are column-normalised in order to reveal any trends with $\Delta \phi_P$. 
Over-plotted are theoretical values of $\delta v_{\parallel}$ and $\delta v_{\perp}$ as a function of $\Delta \phi_P$, which we shall refer to as the `Alfv\'enic lines', that we calculate from Equations \ref{eq:dvp} and \ref{eq:dvop}. The values of $\sqrt{r_A}$ (which effectively controls the amplitudes of the Alfv\'enic lines) are 0.7, 0.6, and 0.5 (corresponding to $r_A \simeq 0.48$, 0.36, and 0.25),  for the $r = 0.39$, 0.65, and \SI{0.90}{AU} distance bins, respectively. 
These values are chosen by inspection to best agree with the data across all values of $\Delta \phi_P$, as minimising  fits to the data is strongly biased towards the high data occurrence near $\Delta \phi_P=0$. These curves thus only serve for illustrative purposes. Note that the chosen values of $r_A$ are not entirely arbitrary, since typically $r_A<1$ in the solar wind  and $r_A$ is observed to decrease with distance \citep{Marsch1990,Bruno2005}.

$\delta v_{\parallel}/v_A$ data follows the general trend illustrated by the Alfv\'enic line; larger deflections of the magnetic field from the Parker spiral direction correspond to greater enhancements in $v_{\parallel}$ relative to the unperturbed value (i.e., the rolling mean). 
This trend is clearest in the two innermost distance bins. 
$\delta v_{\parallel}/v_A$ is roughly bound between a lower value just below zero, and an upper value which traces the chosen Alfv\'enic lines.
In the outermost distance bin, the spread of $\delta v_{\parallel}/v_A$ is generally greater, and a larger portion of the samples have  $\delta v_{\parallel}/v_A$ which falls below the Alfv\'enic line.

Samples for which $|\Delta \phi_P|> \SI{90}{\degree}$ (roughly the inverted HMF cut off) obey similar trends to those where $|\Delta\phi_P|<\SI{90}{\degree}$. The one dimensional histograms of $\delta v_{\parallel}$ (located to the right of each panel) show that $\delta v_{\parallel}$ is approximately centred around zero for the uninverted samples (blue), but is skewed towards positive values for the inverted samples (orange) as a result of the trend for increasing $\delta v_{\parallel}/v_A$ with increasing $|\Delta\phi_P|$.

$\delta v_{\perp}/v_A$ also exhibits trends with $\Delta\phi_P$. 
Most notably there tends to be $\delta v_{\perp}/v_A>0$ ($<0$) for $\Delta \phi_P >0$ ($\Delta \phi_P <0$). This is clearest with the least spread for the innermost distance bin.
The spread of $\delta v_{\perp}/v_A$  about the Alfv\'enic lines appears wider than for  $\Delta \phi_P$, and in each sector the Alfv\'enic line does not appear to be as effective an upper bound (although note these histograms have different $y$-axis scales). There is also greater asymmetry in the positive and negative $\Delta \phi_P$ sectors of the histograms.

\begin{figure*}
    \centering
    \includegraphics[width=.7\textwidth]{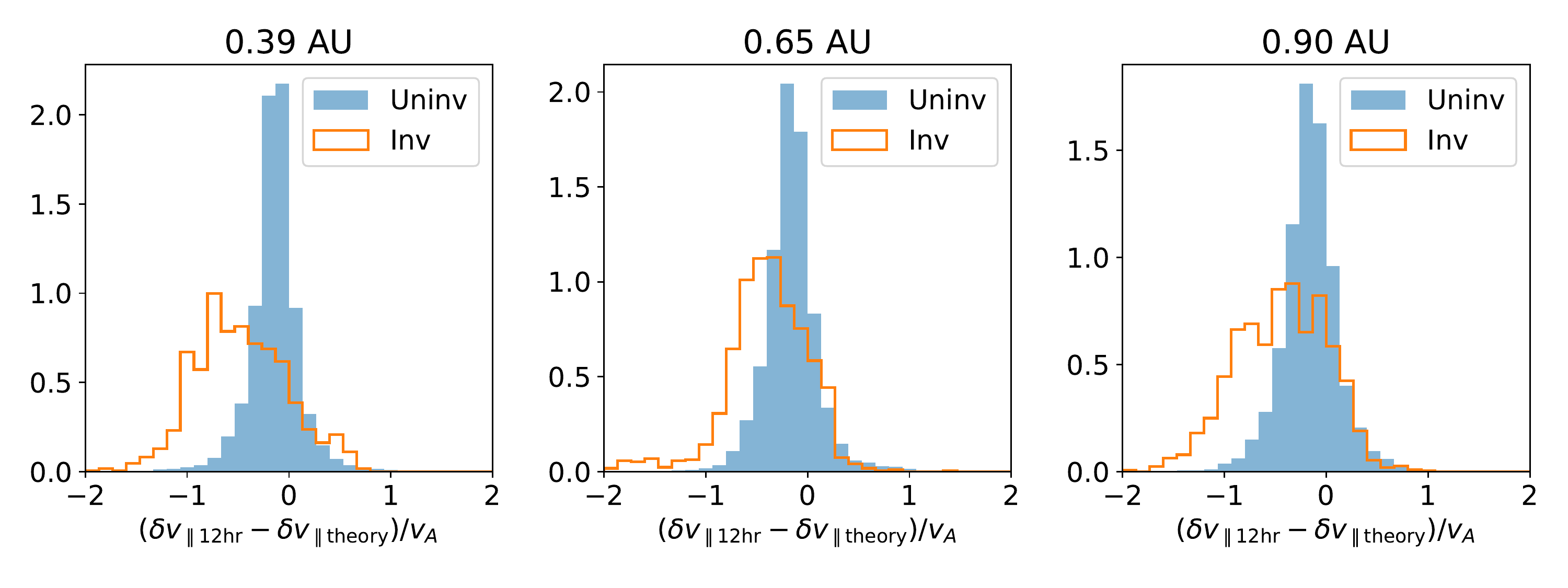}
    \caption{Histograms of the difference between $\delta v_{\parallel}$ derived from the \SI{12}{hr} average, and the theoretical value from Equation \ref{eq:dvp}, $\delta v_{\mathrm{\parallel theory}}$, for uninverted HMF (blue) and inverted HMF (orange) samples, for 3 heliocentric distances. 
    $\delta v_{\mathrm{\parallel theory}}$ is computed using $\sqrt{r_A} = [0.7,0.6,0.5]$ for the three distance bins, in order of increasing $r$.
    Histograms are normalised such that they integrate to unity.}
    \label{fig:dv_error}
\end{figure*}
For each distance bin, we plot in Figure \ref{fig:dv_error}  histograms of  the difference between the measured values of $\delta v_{\parallel}$ (again with \SI{12}{hr} averaging in Equation \ref{eq:dX}) and the respective chosen Alfv\'enic lines in Figure \ref{fig:v_heatmap} ($\delta v_{\mathrm{\parallel theory}}$; the theoretical result of Equation \ref{eq:dvp} for the appropriate value of $\Delta \phi_P$ and prescribed $r_A$). This serves for a qualitative analysis, since our chosen values of $r_A$ can not be correct for all samples in each distance bin.

For all distances in Figure \ref{fig:dv_error}, uninverted samples have $(\delta v_{\mathrm{\parallel 12 hr}}-\delta v_{\mathrm{\parallel theory}})/v_A$ concentrated about zero, and skewing slightly negative. This reflects the tendency for samples shown in Figure \ref{fig:v_heatmap} to follow, and fall just below, the Alfv\'enic lines, which for uninverted samples are close to zero.
The distributions  for inverted samples in Figure \ref{fig:dv_error} have a larger negative extension than for the uninverted samples. Despite this, they also extend to roughly the same positive level as for the uninverted samples. 
This result again shows $\delta v_{\mathrm{\parallel 12 hr}}$ to be roughly bound between chosen Alfv\'enic lines and zero.
Shifts in the appearance of these peaks are likely due to a combination of factors, including the individual values of $r_A$ and the distributions of $\Delta \phi_P$ in each distance bin.

\subsection{Separation by Collisional Age}

To test whether background solar wind conditions have a bearing on inversion properties,
proton collisional age, $A_c$, can be used to classify the solar wind as an alternative to solar wind speed. Traditionally `slow' solar wind tends to a higher collisional age \citep[e.g., Figure 2 of][]{Kasper2008} than the `fast' solar wind. 
$A_c$ expresses the number of collisional time scales, $\tau$, which elapse over the course of the solar wind's propagation to the observer:
\begin{equation}
    A_c = \frac{r}{v_r\tau}
\end{equation}
where $v_r$ is the proton radial velocity. This formulation assumes the solar wind speed and the collision time is constant over all distances. 
Here we employ the proton self-collisional age from \cite{Maruca2013}:
\begin{equation}\label{eq:ageform}
    A_c = \left( \SI{1.31e7}{\centi\meter\cubed\kilo\meter\raiseto{3/2}\kelvin\per \second\per\astronomicalunit}\right)\left(\frac{n_p}{v_{r}T_p^{3/2}}\right)r\lambda_p,
\end{equation}
where $T_p$ is the proton temperature obtained from the corefit data set. $\lambda_p$ is the proton coulomb logarithm:
\begin{equation}
    \lambda_p = 9.42 + \ln{\left(\frac{T_p^{3/2}}{n_p^{1/2}}\SI{}{\per\raiseto{3/2}\centi\meter\per\raiseto{3/2}\kelvin}\right)},
\end{equation}
The inclusion of other parameters in calculating $A_c$ allows for an alternative classification to using solar wind speed alone, which has been found at times to not clearly separate solar wind intervals of different solar origin \citep[e.g.,][]{Stansby2020}, and may bias our results since it is an important factor in computing $\delta v_{\parallel}$. Splitting the \textit{Helios} data into further groups in this way may increase the clarity of trends which we infer from e.g., Figure \ref{fig:v_heatmap}, since variable factors should be more consistent for similar solar wind streams.

To employ $A_c$ as an indicator of solar wind type, we choose thresholds with which to split the data based on the lower and upper quartile values of $A_c$ observed in each distance bin. Since we are only comparing $A_c$ within individual bins, we shall refer to the samples which fall below the lower (exceed the upper) quartile value as low (high) `relative' collisional age.
By construction, the low and high relative collisional age samples make up the same proportion of samples in each distance bin.  This ensures a classification which is relatively consistent with distance, since the distribution of solar wind type should not change strongly with $r$ (excluding the effects of interaction regions).

\begin{figure*}
    \centering
    \includegraphics[clip,trim={0 10.05cm 0 0},width=.8\textwidth]{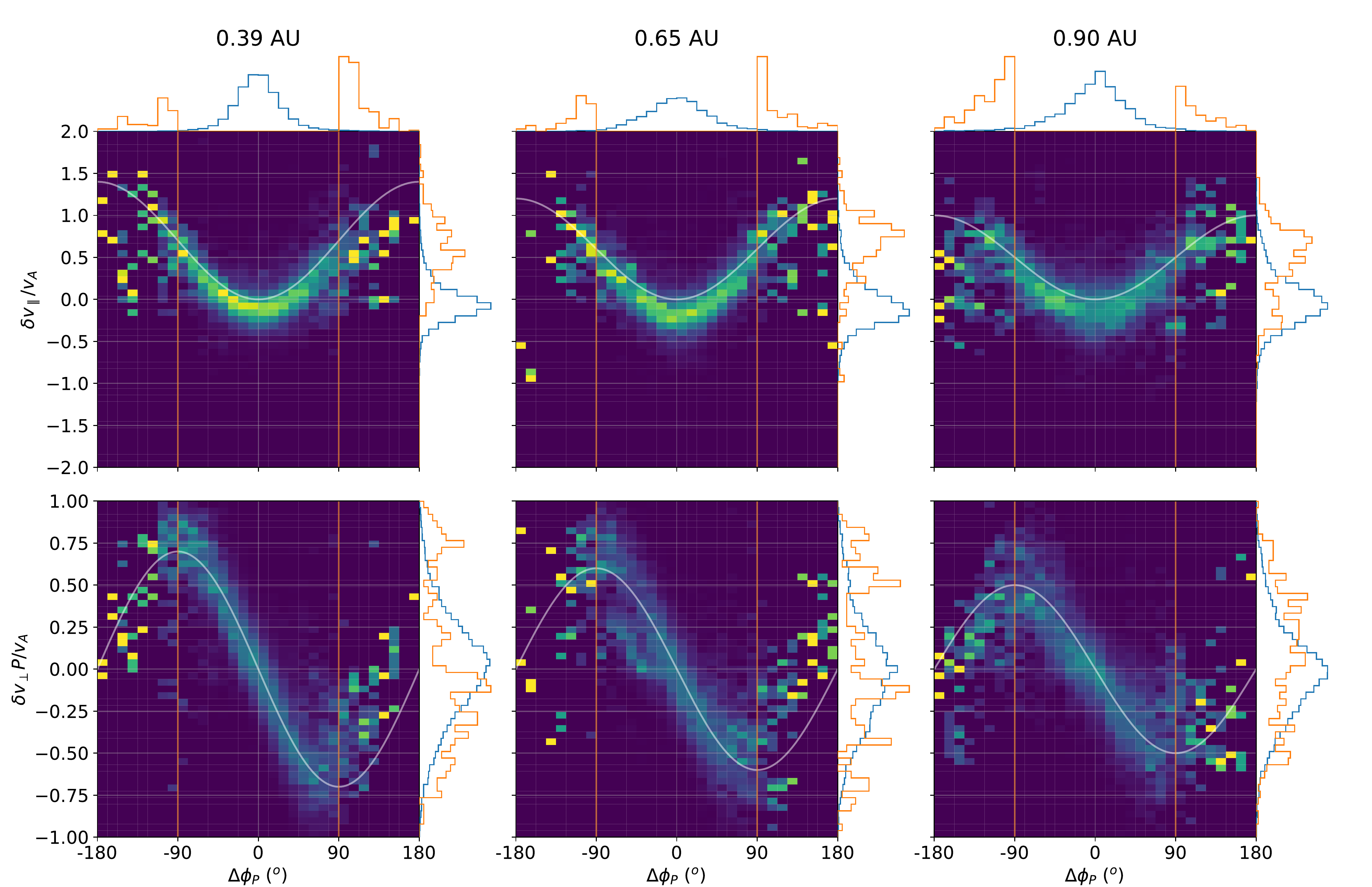}
    \caption{2-dimensional histograms of $ \delta v_{\parallel}/v_A$ and $\delta v_{\perp}/v_A$ against $\Delta \phi_P$ for three radial distance bins in the same format as Figure \ref{fig:v_heatmap}. The data only include samples within the lowest quartile of $A_c$ for the relevant distance bin. Above the top row of panels, the blue histogram shows the distribution of $\Delta \phi_P$ for the  samples included in the relevant 2-dimensional histograms. The orange histogram shows the distribution  of $\Delta \phi_P$ for the included inverted HMF samples only. The orange and blue histograms are normalised independently.
    To the right of each panel, normalised histograms show the distribution of that panel's $y$-axis parameter for  uninverted and inverted HMF samples combined (blue line) and for inverted samples only (orange line). }
    \label{fig:vmap_young}
\end{figure*}
\begin{figure*}
    \centering
    \includegraphics[clip,trim={0 10.05cm 0 0},width=.8\textwidth]{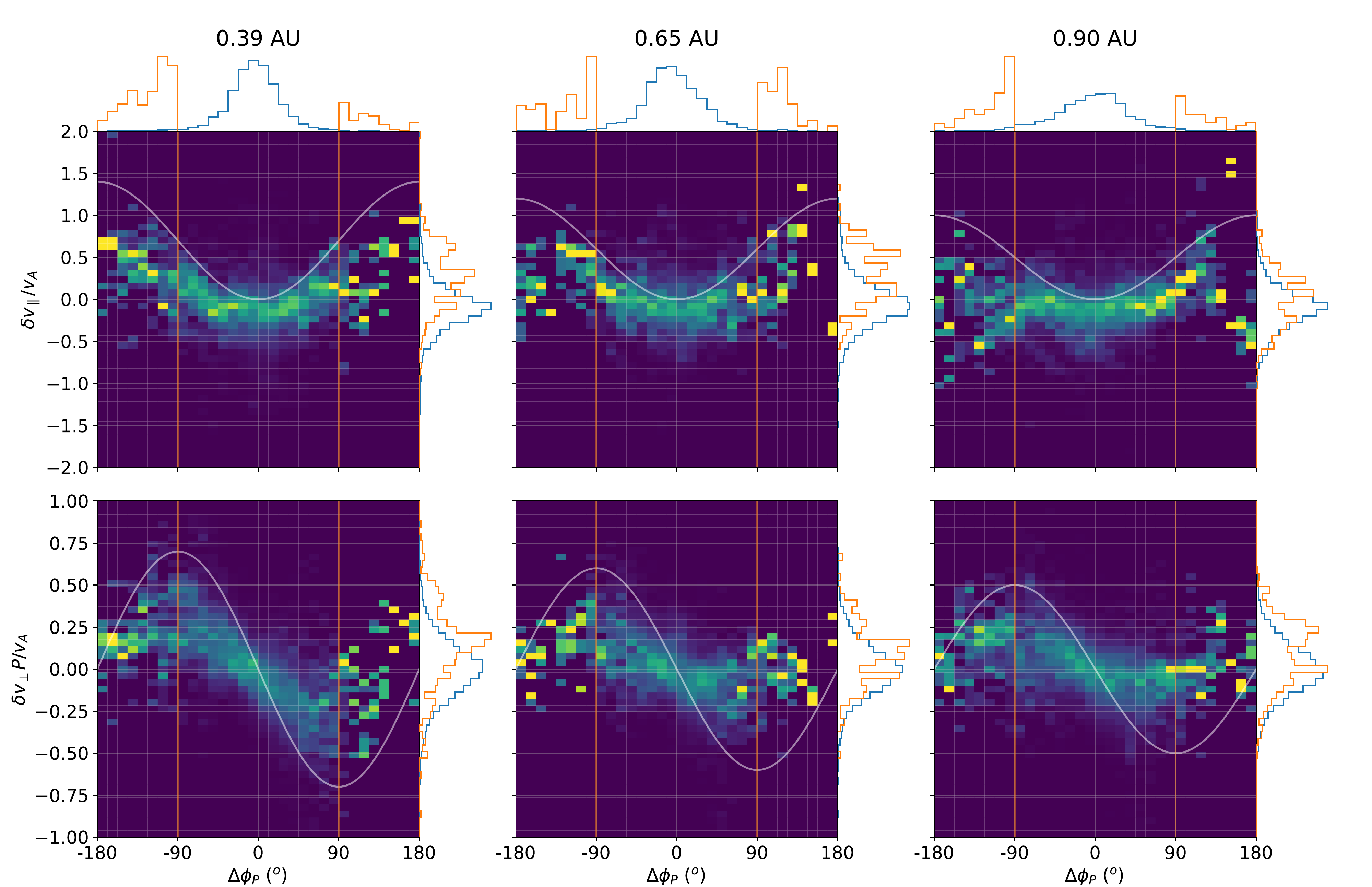}
    \caption{2-dimensional histograms of $ \delta v_{\parallel}/v_A$ and $\delta v_{\perp}/v_A$ against $\Delta \phi_P$ for three radial distance bins in the same format as Figure \ref{fig:v_heatmap}. The data only include samples within the highest quartile of $A_c$ for the relevant distance bin. 
    Above the top row of panels, the blue histogram shows the distribution of $\Delta \phi_P$ for the  samples included in the relevant 2-dimensional histograms. The orange histogram shows the distribution  of $\Delta \phi_P$ for the included inverted HMF samples only. The orange and blue histograms are normalised independently. 
    To the right of each panel, normalised histograms show the distribution of that panel's $y$-axis parameter for  uninverted and inverted HMF samples combined (blue line) and for inverted samples only (orange line). }
    \label{fig:vmap_old}
\end{figure*}

Figure \ref{fig:vmap_young} displays histograms of $\delta v_{\parallel}/v_A$ and $\delta v_{\perp}/v_A$ similar to Figure \ref{fig:v_heatmap}, but for only those solar wind samples with low relative $A_c$, per the above definition. We again plot the Alfv\'enic lines from Equations \ref{eq:dvp} and \ref{eq:dvop}, using the same values of $r_A$. The histograms of $\Delta \phi_P$ show a deal of asymmetry for inverted and uninverted flux, which is probably a result of the reduction in number of samples due to further splitting the data.
Overall, $\delta v_{\parallel}/v_A$ and $\delta v_{\perp}/v_A$ data in Figure \ref{fig:vmap_young} track the Alfv\'enic lines remarkably  closely in comparison to the data set as a whole, shown in  Figure \ref{fig:v_heatmap}. The 1-dimensional histograms of $\delta v_{\parallel}/v_A$ show that very few samples fall near $\delta v_{\parallel}=0$ for inverted HMF. 
While in all bins the Alfv\'enic relationship is clear, $\delta v_{\parallel}/v_A$ and $\delta v_{\perp}/v_A$ both become more spread out with increasing distance from the Sun.

Figure \ref{fig:vmap_old} follows the same format as Figure \ref{fig:vmap_young} but for high relative $A_c$ values in each distance bin.  In general $\delta v_{\parallel}$ is more spread out at all angles, but with some correlation with $\Delta \phi_P$ found at the larger angles.
The data here do not follow the Alfv\'enic lines as closely as the full data in Figure \ref{fig:v_heatmap} or the low relative $A_c$ data in Figure \ref{fig:vmap_young}; $\delta v_{\parallel}$ and $\delta v_{\perp}$ both more commonly fall below. This may be related to the lower $r_A$ which is characteristic of the slow solar wind \citep{Marsch1990,Bruno2005}.
Additionally, more samples are found around $\delta v_{\parallel}=0$ than for the low relative $A_c$ data.
$\delta v_{\perp}$, however, is still predominantly $>0$ ($<0$) for $\Delta \phi_P<0 $ ($>0$). 
We also note that the number of inverted HMF samples in the high relative $A_c$ quartile of data is in all bins greater than the number in the low relative $A_c$ quartile.

\subsection{Variations in Density and Field Strength}

\begin{figure*}
    \centering
    \includegraphics[width=.8\textwidth]{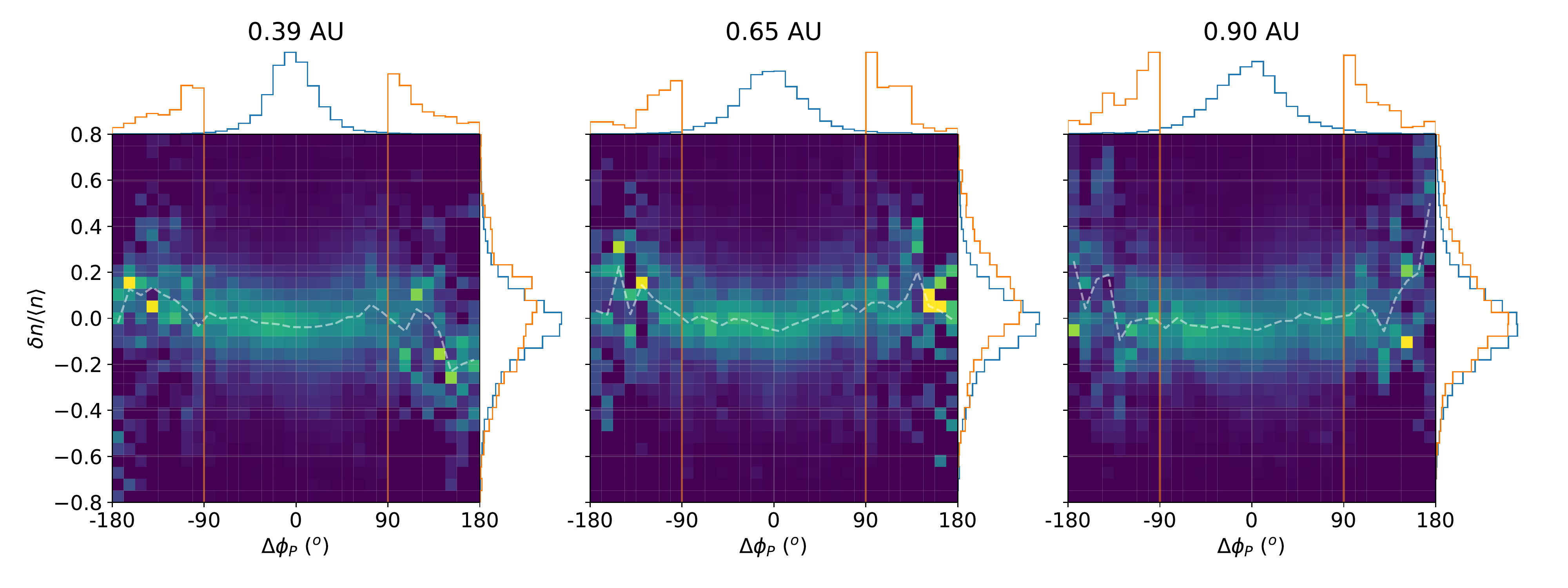}
    \caption{2-dimensional histograms of $\delta n/\langle n\rangle = (n - \langle n\rangle)/\langle n\rangle$ against $\Delta \phi_P$ for three radial distance bins. $\langle n\rangle$ is calculated over \SI{12}{hours}.  A dashed white line shows the median value in each column of  $\Delta \phi_P$.  The figure is in the same format as the top row of Figure \ref{fig:v_heatmap}. 
    Above each panel, the blue histogram shows the distribution of $\Delta \phi_P$ for all samples in a given distance bin, while the orange histogram shows the distribution  of $\Delta \phi_P$ for  inverted HMF samples only. The orange and blue histograms are normalised independently.  To the right of each panel, normalised histograms show the distribution of $\delta n/\langle n\rangle$ for combined uninverted and inverted HMF samples  (blue line) and for inverted samples only (orange line).}
    \label{fig:n_heatmap}
\end{figure*}
We now explore the changes in plasma density and magnetic field intensity associated with these inversions.
Figure \ref{fig:n_heatmap} plots $\delta n/\langle n\rangle$, the normalised difference between proton density for a given sample and the \SI{12}{hr} running average calculated per Equation \ref{eq:dX}, against $\Delta\phi_P$, in the same form as the first row of Figure \ref{fig:v_heatmap}. 
Generally $\delta n/\langle n\rangle$ exhibits greater spread at larger $|\Delta\phi_P|$. 
The one dimensional histograms to the side of each plot show that 
$\delta n/\langle n\rangle$ for uninverted HMF is approximately centred on zero. Conversely, for inverted HMF $\delta n/\langle n\rangle$ is skewed slightly positive. 
Inverted HMF samples thus tend to have slightly larger changes in density than uninverted samples, and these changes in density tend towards being increases. One exception to this is at $\Delta \phi_P\gtrsim\SI{120}{\degree}$ in the inner distance bin, where $\delta n/\langle n\rangle$ extends to negative values. 
Due to large outliers, we compute root median square (RMedS) values of $\delta n/\langle n\rangle$, to compare the characteristic size of $\delta n /\langle n\rangle$ for inverted and uninverted data.
RMedS values for uninverted (inverted) HMF samples are in the ranges 0.21--0.22 (0.25--0.29) for each radial bin, confirming the slightly larger changes in density associated with larger HMF deflections.   

\begin{figure*}
    \centering
    \includegraphics[width=.8\textwidth]{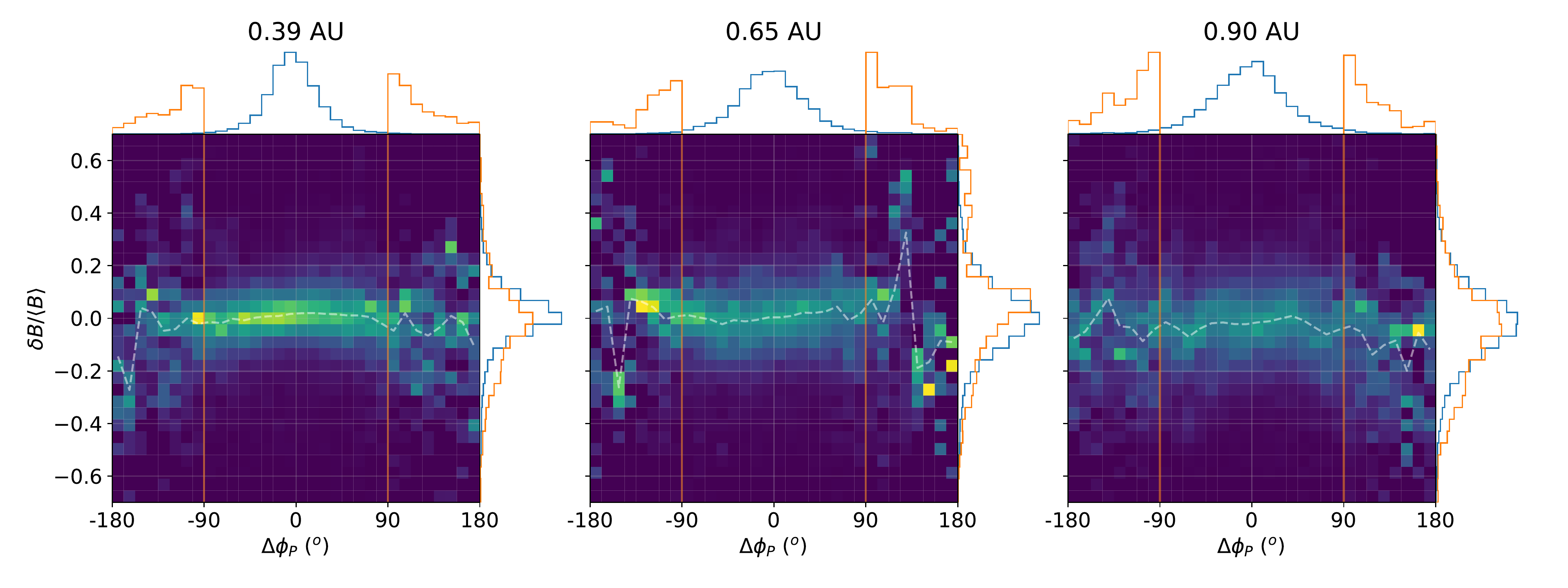}
    \caption{2-dimensional histograms of $\delta B/\langle B\rangle = (B - \langle B\rangle)/\langle B\rangle$ against $\Delta \phi_P$ for three radial distance bins. $\langle B\rangle$ is calculated over \SI{12}{hours}. A dashed white line shows the median value in each column of  $\Delta \phi_P$. 
    The figure is in the same format as the top row of Figure \ref{fig:v_heatmap}. 
    Above each panel, the blue histogram shows the distribution of $\Delta \phi_P$ for all samples in a given distance bin, while the orange histogram shows the distribution  of $\Delta \phi_P$ for  inverted HMF samples only. The orange and blue histograms are normalised independently.  To the right of each panel, normalised histograms show the distribution of $\delta B/\langle B\rangle$ for uninverted and inverted HMF samples combined  (blue line) and for inverted samples only (orange line).}
    \label{fig:B_heatmap}
\end{figure*}

Figure \ref{fig:B_heatmap} plots histograms of $\delta B/\langle B\rangle$, as calculated using Equation \ref{eq:dX}, against $\Delta \phi_P$, in the same format as Figure \ref{fig:n_heatmap}. ($B$ in non-boldface here represents the magnitude of the magnetic field vector: $B=|\mathbf{B}|$.) $\delta B/\langle B\rangle$  is more broadly spread for inverted than uninverted HMF samples. For uninverted samples  $\delta B/\langle B\rangle$ is centred around zero, while for inverted samples it skews towards negative values. Note that $\delta B/\langle B\rangle$  for inverted samples in the central distance bin is distributed quite irregularly, with values at positive and negative extremes for the $\Delta \phi>\SI{90}{\degree}$ sector.
Weak decreases in $B$ during inversions were observed in several of the switchbacks observed by PSP \citep[Figure 2a of][]{Bale2019}.
RMedS  values of $\delta B/\langle B\rangle$ in uninverted (inverted) HMF samples range 0.06--0.09 (0.11--0.13).
The typical change in magnetic field magnitude during inversions is thus slightly larger than that at other times.

\begin{figure*}
    \centering
    \includegraphics[clip,trim={0 7cm 0 0},width=.8\textwidth]{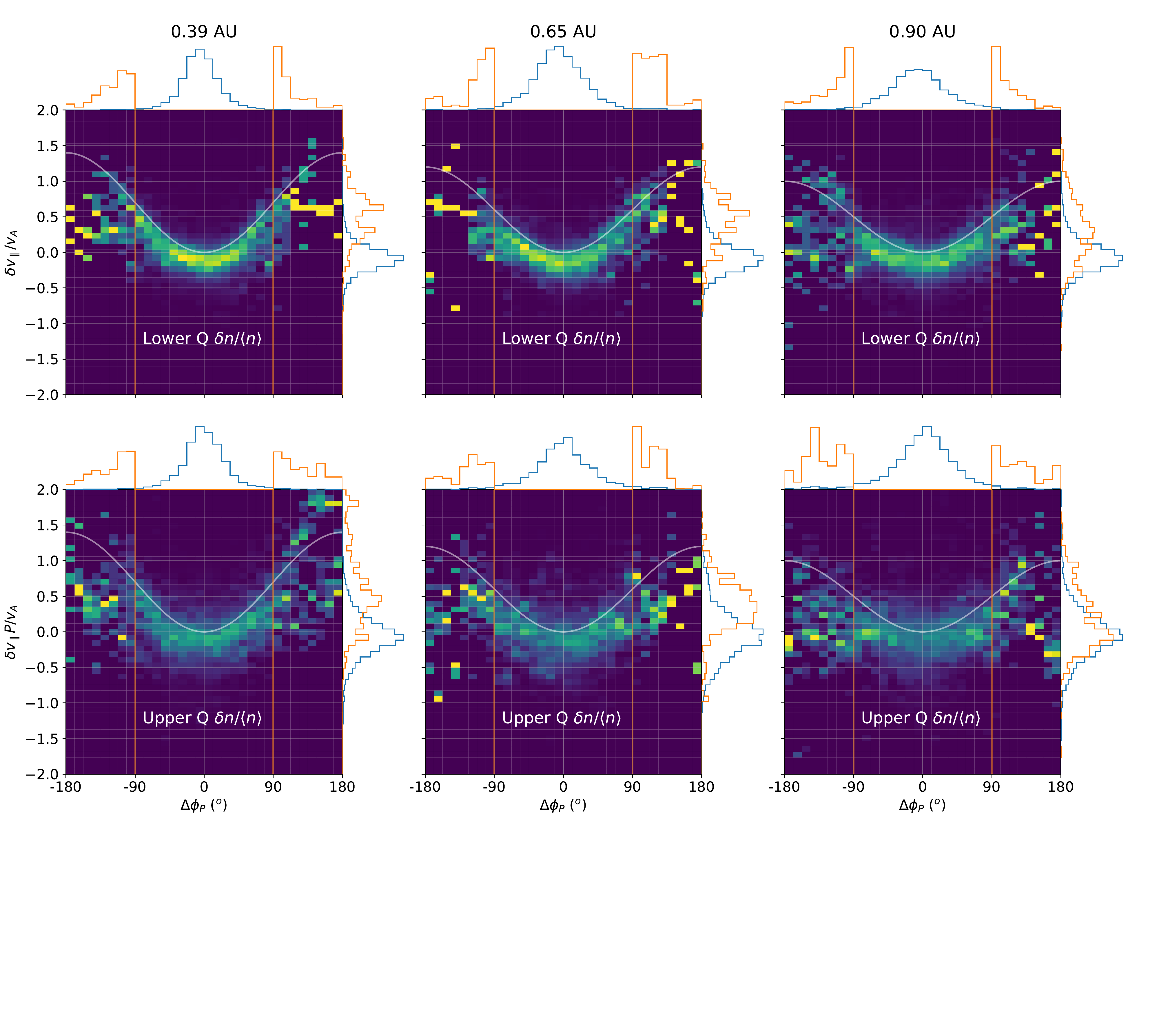}
    \caption{2-dimensional histograms of $ \delta v_{\parallel}/v_A$ against $\Delta \phi_P$ for three radial distance bins in the same format as the top row of Figure \ref{fig:v_heatmap}. The top (bottom) row of plots include data within the lower (upper) quartile of $|\delta n/\langle n\rangle|$ within each distance bin.
    Above each panel, the blue histogram shows the distribution of $\Delta \phi_P$ for all samples included in the respective 2-dimensional histogram, while the orange histogram shows the distribution  of $\Delta \phi_P$ for included inverted HMF samples only. The orange and blue histograms are normalised independently.  To the right of each panel, normalised histograms show the distribution of $ \delta v_{\parallel}/v_A$ for combined uninverted and inverted HMF samples  (blue line) and for inverted samples only (orange line) included in the respective 2-dimensional histogram.}
    \label{fig:v_heatmap_nsplit}
\end{figure*}

\begin{figure*}
    \centering
    \includegraphics[clip,trim={0 7cm 0 0},width=.8\textwidth]{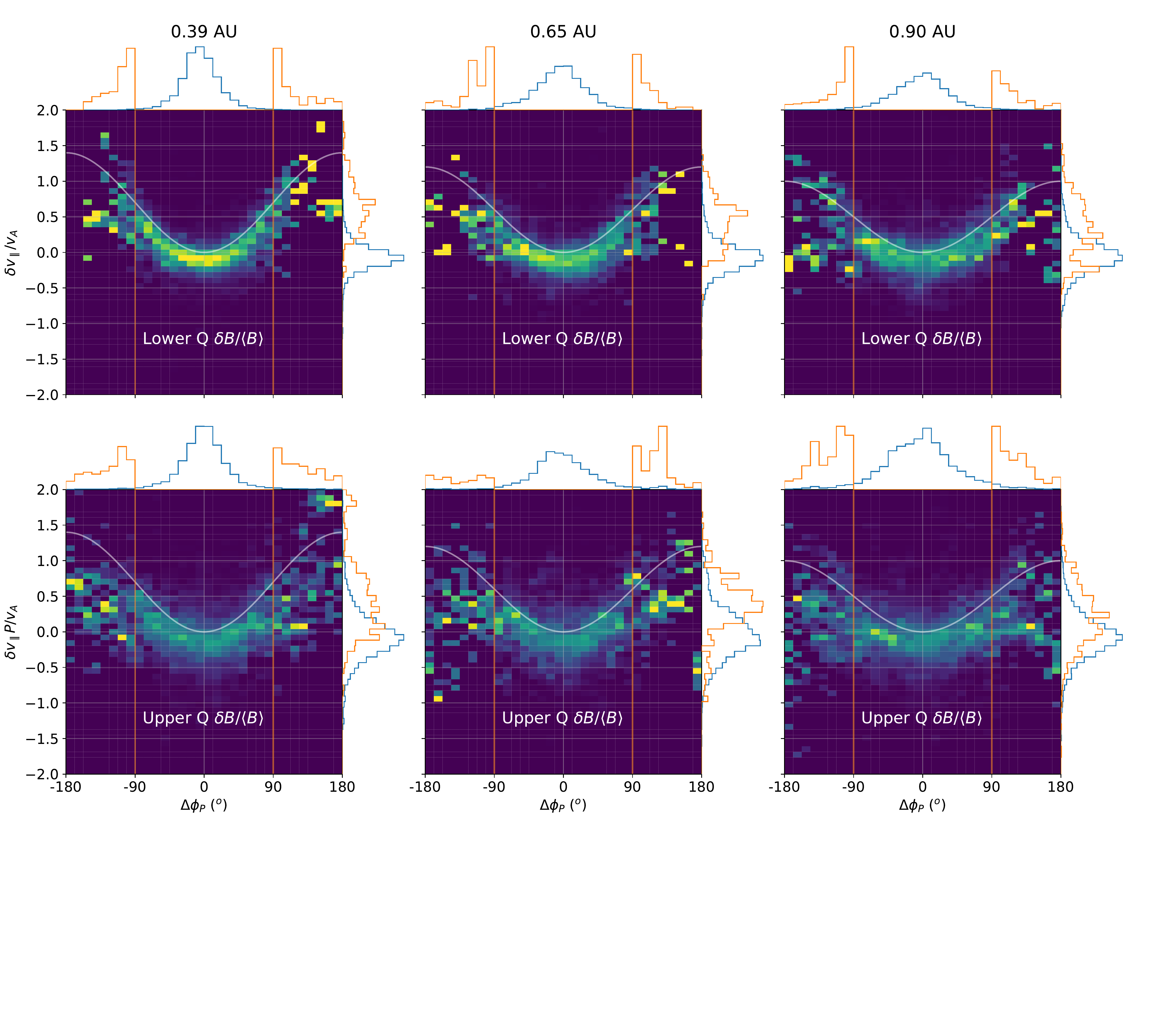}
    \caption{2-dimensional histograms of $ \delta v_{\parallel}/v_A$ against $\Delta \phi_P$ for three radial distance bins in the same format as the top row of Figure \ref{fig:v_heatmap}. The top (bottom) row of plots include data within the lower (upper) quartile of $|\delta B/\langle B\rangle|$ within each distance bin.
    Above each panel, the blue histogram shows the distribution of $\Delta \phi_P$ for all samples included in the respective 2-dimensional histogram, while the orange histogram shows the distribution  of $\Delta \phi_P$ for included inverted HMF samples only. The orange and blue histograms are normalised independently.  To the right of each panel, normalised histograms show the distribution of $ \delta v_{\parallel}/v_A$ for  uninverted and inverted HMF samples combined (blue line) and for inverted samples only (orange line) included in the respective 2-dimensional histogram.
    }
    \label{fig:v_heatmap_Bsplit}
\end{figure*}

We further investigate the association between $\delta n/\langle n\rangle$ and $\delta B/\langle B\rangle$, and the Alfv\'enic  $\delta \mathbf{v}$-$\delta \mathbf{B}$ relationships reported above. To do so we split the data in each distance bin into upper and lower quartiles of $|\delta n/\langle n\rangle|$ and $|\delta B/\langle B\rangle|$, and reproduce 
the $\delta v_{\parallel}/v_A$ histograms of Figure \ref{fig:v_heatmap} for each subset of data. These are shown for $|\delta n/\langle n\rangle|$ in Figure \ref{fig:v_heatmap_nsplit} and $|\delta B/\langle B\rangle|$ in Figure \ref{fig:v_heatmap_Bsplit}.  
The sets of histograms generated using the data within the lower quartiles of $|\delta n/\langle n\rangle|$ and $|\delta B/\langle B\rangle|$ (top rows of each figure) generally follow the illustrative Alfv\'enic lines reasonably closely. The inner distance bin for low $|\delta B/\langle B\rangle|$ exhibits the Alfv\'enic relationship particularly clearly. 
The $\delta v_{\parallel}/v_A$-$\Delta\phi_P$ relationships here are in general quite similar to the low relative $A_c$ data. 

The histograms generated using the data within the upper quartiles of $\delta n/\langle n\rangle$ and $\delta B/\langle B\rangle$ (bottom rows of each figure) have a far greater spread, mostly below the Alfv\'enic lines. However, there is still a visible trend of increasing $\delta v_{\parallel}/v_A$ with increasing $|\Delta \phi_P|$. 
In the outer distance bin, the data within the upper quartiles of $|\delta n/\langle n\rangle|$ and $|\delta B/\langle B\rangle|$ values (bottom-right histograms of Figures \ref{fig:v_heatmap_nsplit}) show the weakest trends. The 1-dimensional histograms of $\delta v_{\parallel}$  on the right hand side of these panels show that the inverted HMF samples have a considerable component which falls near $\delta v_{\parallel}=0$.
These results show differences in the $\delta \mathbf{v}$-$\delta \mathbf{B}$  relationship are associated with $\delta n/\langle n\rangle$ (and $\delta B/\langle B\rangle$), or the compressibility associated with the fluctuation. 
There is possible weak bias in this analysis, since larger $|\delta n/\langle n\rangle|$ and $|\delta B/\langle B\rangle|$ are associated with slightly greater $\Delta \phi_P$ for inversions (as shown by the 1-dimensional $\Delta \phi_P$ histograms above each panel).

\section{Discussion}\label{sec:disc}

\subsection{Limitations and Caveats for Results}\label{sub:caveats}

Several factors are important for a nuanced interpretation of this study's results. 
First, 
because inversions exist on a range of scales \citep{Matteini2014,Owens2020,TDW2020} there is no one ideal time scale for this analysis. Thus our results here reflect the available minimum \SI{40}{s} time resolution and chosen averaging times in Equations \ref{eq:dpm}  and  \ref{eq:dX}.
We have produced Figure \ref{fig:v_heatmap}, from which we first recovered $\delta \mathbf{v}$-$\delta \mathbf{B}$  relationships,  using  1, 4, 12, and \SI{24}{hr} averages for the background, and find that 12 hours best exhibits the Alfv\'enic trend with least spread, although it is present for all time scales.  

Due to field of view limitations of the \textit{Helios} E1 instrument we impose a condition in Section \ref{sec:datmeth} \citep[detailed in][]{Macneil2020sunward} which excludes HMF samples with large non-azimuthal field components. A consequence of this is that only those large field deflections which are aligned roughly within the $R$-$T$ plane are considered in this study. Our results thus strictly only apply for inversions which occur through azimuthal field deflections. This aspect of the analysis likely contributes  to the clarity of the Alfv\'enic trend as seen in e.g., Figure \ref{fig:v_heatmap}, since the derivation of Equation \ref{eq:dvp} assumes that the field rotates only in the azimuthal plane.

$\Delta \phi_P$ represents the magnetic field azimuthal deflection angle away from the ideal Parker spiral direction. Analysis  using $\Delta \phi_P$ assumes that the unperturbed magnetic field direction, $\phi_0$, is well-represented by the Parker spiral angle $\phi_P$. Employing rolling averages to instead estimate the background field would be problematic for this study, since local magnetic inversions strongly perturb the field and will influence such averages.
Offsets between $\phi_0$ and $\phi_P$ will lead to an additional spread in $\delta v_{\parallel (\perp)}$ for an otherwise ideal Alfv\'enic fluctuation, as demonstrated through modelling in Appendix \ref{app:angles}.
If the errors in field angle are statistically symmetric, this will lead to a spread in  $\delta v_{\parallel}$ which extends more below the true Alfv\'enic relationship than above. 
Meanwhile, the error in $\delta v_{\perp}$ would manifest in a more even spread above and below the Alfv\'enic line, but which is greater for larger field deflections. 
Both of these are roughly compatible with the observed spreads in $\delta v_{\parallel (\perp)}$. Conclusions regarding the overall spread in these parameters must therefore recognise this contribution.

Section \ref{sec:datmeth} details the additional step taken in classifying inverted HMF, in which we discard samples which appear to be inverted based on electron PADs, but do not possess the non-dominant field polarity in a rolling \SI{12}{hr} window. This removes a sub-population of inversions which were possibly mis-identified based on strahl alone.
As highlighted in Appendix \ref{app:correction}, this procedure will also remove inversions which occur near a sector boundary, occur in bursts, or are large in size.
Removing these inversions may have a knock-on effect for our results, since they would all be likely to appear non-Alfv\'enic when comparing e.g., $\delta v_{\parallel}$ to $\Delta \phi_P$. This is because of difficulties in estimating an accurate background velocity under these conditions, in addition to the possibility of such inversions, particularly those near the HCS, being non-Alfv\'enic. Large scale inverted HMF which is embedded in the solar wind and so does not produce a local deflection in the field \citep[e.g.,][]{Crooker2004} will also be removed by this method. 
Thus, the conclusions drawn below only apply for inversions which take the form of clearly-defined, transient, field deflections.
This likely contributes to the clarity of the relationships found in the previous section.

\subsection{$\delta \mathbf{v}$-$\delta \mathbf{B}$ Correlation During Magnetic Inversions}

Results in Figure \ref{fig:v_heatmap} show that azimuthal departures of the field from the Parker spiral direction, including large deflections which constitute HMF inversions,  are often accompanied by increases in the Parker spiral velocity component $v_{\parallel}$.
The magnitude of the $v_{\parallel}$ increase is such that $\delta v_{\parallel}$ is roughly bound between zero and the chosen Alfv\'enic lines. Thus, at greater $|\Delta\phi_P|$, larger $v_{\parallel}$ enhancements are accessible.
This, and the observation that $\delta v_{\perp}$ rarely has the opposite sign to the illustrative Alfv\'enic lines, indicates that inversions studied here commonly propagate anti-sunward along the background field, and are Alfv\'enic in nature. 
For the purpose of this discussion section, we shall consider inversions to be `Alfv\'enic' if the deflection which inverts the field is coincident with an anti-correlated change in the plasma velocity, i.e., as would be shown in Figure \ref{fig:v_heatmap}.
Alfv\'enic inversions have been observed prior in the inner portion of the \textit{Helios} 2 orbit by \cite{Horbury2018}.
Here we find clear evidence of Alfv\'enic inversions, particularly in the innermost distance bin, but also over the full \textit{Helios} distance range, including several years of data, and for a range of solar wind types.
These results agree with the correlation between field azimuthal angle and velocity enhancement in Ulysses measurements of polar coronal hole wind $> \SI{1}{AU}$,  as reported by \cite{Matteini2014}, who also demonstrated their Alfv\'enic nature. %
Figure \ref{fig:dvpm} shows that inversions most commonly exhibit a `spike' in their $v_{\parallel}$ profile, in contrast with the overall data which mostly exhibit a decreasing profile. This is also consistent with a majority of these samples being Alfv\'enic switchbacks, where the reversal in the field is simultaneous with a short-lived velocity enhancement \citep{Horbury2018,Bale2019,Kasper2019}. These spikes are revealed in hourly-averaged $v_{\parallel}$ differences, so Alfv\'enic profiles of much longer or shorter duration inversions may be missed in our analysis.

We consider an `ideal' Alfv\'enic fluctuation to have $r_A=1$, indicating energy balance between the magnetic field and plasma components of the fluctuation. The illustrative Alfv\'enic lines on Figure \ref{fig:v_heatmap} have $r_A<1$, indicating that the energy of such a fluctuation is weighted towards the magnetic field rather than the plasma, as is typical of the solar wind \citep{Marsch1990,Bruno2005}. 
A given sample falling below the Alfv\'enic line, as most inverted HMF samples do in Figure \ref{fig:v_heatmap}, indicates that even less of the energy in the fluctuation is contained in the plasma, and thus it is further from the ideal Alfv\'enic case. This may be related to the properties of the surrounding plasma, or the fluctuation itself. Independently measuring $r_A$ to verify this is problematic however, because it requires quantifying $\delta \mathbf{v}$ and $\delta\mathbf{B}$, which is in essence the analysis already carried out here.
Many inverted HMF samples, particularly in the outermost distance bin, have $\delta v_{\parallel}$ and/or $\delta v_{\perp}$ very close to, or just below, zero. For an Alfv\'enic fluctuation, this would require extremely small $r_A$, since Equations \ref{eq:dvp} and \ref{eq:dvop} depend on $\sqrt{r_A}$. The spread in estimated $\delta v_{\parallel}$ due to offsets between $\phi_0$ and $\phi_P$ may be partially responsible, but this is unlikely at large $\Delta\phi_P$.
Thus there are some inversions here for which an Alfv\'enic description is probably inappropriate (non-Alfv\'enic). However, there is no clear cutoff in $r_A$ below which we can consider the inversion as clearly non-Alfv\'enic.
These inversions thus exist as fluctuations in a continuum between non-Alfv\'enic and Alfv\'enic (albeit with a dominant magnetic field component).

A reasonable upper bound on $\delta v_{\parallel(\perp)}$ is established using Alfv\'enic lines with an $r_A$ which drops off with $r$ (Figure \ref{fig:dv_error} as well as e.g., Figure \ref{fig:v_heatmap_Bsplit}).
Thus, the upper limit of the portion of energy contained in the plasma component of these deflections decreases with distance, in a roughly linear fashion. (A more dedicated study of the variation in $r_A$ would be better suited to quantify this drop-off.) In this way the inversions depart further from the ideal Alfv\'enic case with $r$.
Decreasing overall Alfv\'enicity with increasing $r$ has been observed prior over the solar distances considered in this study \citep[e.g.,][]{Roberts1987,Bruno2007}, and appears to be consistent with this observation. 
If inversions are being formed locally, then the upper bound on $\delta v_{\parallel}$ is consistent with their forming with energy partitioning consistent with local $r_A$. If they are otherwise formed at the Sun, then  the velocity component of the fluctuation decays more rapidly than that in the field component, in line with $r_A$ for the solar wind overall.
In the same data set as the present study, \cite{Macneil2020inv} observed the greatest fraction of inverted HMF samples near \SI{1}{AU}, where we find inversions to also be the furthest from ideally Alfv\'enic (particularly for high relative collisional age solar wind).
Thus, while inversions commonly follow something akin to an Alfv\'enic relationship with $\Delta \phi_P$, strong Alfv\'enicity is not a prerequisite for inversions to exist.

Considering samples with a low collisional age ($A_c$) relative to the rest of the distance bin reveals a far clearer Alfv\'enic relationship than is found for  the data set as a whole. The grouping around the chosen Alfv\'enic lines for the inner two distance bins in Figure \ref{fig:vmap_young} indicates a consistent $r_A$. Few inverted HMF samples fall near $\delta v_{\parallel}=0$, so an Alfv\'enic description is appropriate for nearly all of the studied inversions with low relative collisional age. This could simply be related to the general Alfv\'enic nature of the fast solar wind, which low $A_c$ proxies for. The increased spread in $\delta v_{\parallel(\perp)}$ furthest from the Sun may be a result of decreasing Alfv\'enicity, or perhaps increased offsets between $\phi_P$ and $\phi_0$.

Meanwhile, high relative $A_c$ samples are far more spread out in $\delta v_{\parallel}$, and fall short of the same Alfv\'enic lines. Particularly in the outer distance bin, many large field deflections fall near $\delta v_{\parallel}=0$, indicating that Alfv\'enic character is very weak or absent for such inversions. This is consistent with the low Alfv\'enicity associated with the slow solar wind. It could also be related to a greater degree of solar wind processing, if departures from ideal Alfv\'enic properties are here a result of transit effects. 
Interestingly, there are more inverted HMF samples with high than low relative collisional age. This further cements that clear Alfv\'enic signatures are not a requirement for inverted HMF to exist.

Solar wind density ($n$) and magnetic field intensity ($B$) do not change greatly during magnetic inversions, indicating that most of these inversions are at most only weakly compressible. 
The tendency for a slight decrease in $B$ during some PSP-observed inversions has been reported previously by \cite{Bale2019}. The weak increase in $n$ may be related to pressure balance structures.
Weak compressibility is compatible overall with the established properties of PSP-observed switchbacks \citep[][]{Bale2019,Horbury2020}. From Figures \ref{fig:v_heatmap_nsplit} and \ref{fig:v_heatmap_Bsplit}, the spread in $\delta v_{\parallel(\perp)}$ is greater at larger $r$ and for larger changes in $n$ and $B$, such that $\delta v_{\parallel}$ in the outermost distance bin, for large changes in $n$, shows very little enhancement at larger $\Delta \phi_P$. 
This suggests that although most inversions studied here have low compressibility or are incompressible, higher relative compressibility is associated with departures from ideal Alfv\'enic signatures. 
The observations of less Alfv\'enic inversions occurring in high relative collisional age solar wind are likely linked to this result, since high collisional age is indicative of typical `slow' solar wind, which typically possesses more compressive features. 
A very clear Alfv\'enic relationship emerges in Figures \ref{fig:v_heatmap_nsplit} and \ref{fig:v_heatmap_Bsplit} when small changes in $n$, and particularly $B$, accompany the inversion. This is intuitive from the derivation of the Alfv\'enic $\delta v_{\parallel{\perp}}$ relationships in Equations \ref{eq:dvp} and \ref{eq:dvop}, where constant $B$ is an explicit assumption.

Greater departures from ideal Alfv\'enicity for inversions with greater $\delta n$ and $\delta B$, and high relative collisional age, could be  explained on one hand by exposure to compressive structures or fluctuations in the solar wind  eroding the characteristic Alfv\'enic $\delta  \mathbf{v}$-$\delta \mathbf{B}$ correlation.
Such processing, specifically through magnetic compressibility, has been suggested to be responsible for reduction in solar wind Alfv\'enicity as a function of $r$ by \cite{Bruno1991,Damicis2015}. 
Under this framework, Alfv\'enic and non-Alfv\'enic inversions could both be formed through similar processes, such as through interchange reconnection at the Sun, or growing turbulent fluctuations in the heliosphere \citep[as favoured by previous analysis of this data set][]{Macneil2020inv}.
How close to ideally Alfv\'enic a given HMF inversion is  would then be a result of its evolution, rather than formation, and linked to the processing within the surrounding plasma.

An alternative explanation is that non-Alfv\'enic inversions which tend to be more compressible are in fact formed through a different process to the more Alfv\'enic inversions. 
Suitable candidate processes for creating non-Alfv\'enic inversions would be the draping of the HMF over blobs, or inversion formation in regions of solar wind shear, where gradients in density and field strength are more likely present (see Section \ref{sec:intro}).

Of the two above options to explain discrepancies in inversion Alfv\'enicity, the results of this study lead us to favour the first; that variable Alfv\'enic signatures of inversions are a result of different solar wind processing, likely related to the presence of compressive structures or fluctuations.
A key reason for this conclusion is that we find no evidence to suggest that $\delta v_{\parallel}$ studied here does not exist in a continuum between $\delta v_{\parallel}\leq0$ (no Alfv\'enic signatures) and the chosen Alfv\'enic lines, which represent a non-ideal Alfv\'enic relationship. 
The gradual decrease in $r_A$ (both around the upper boundary of $\delta v_{\parallel}$ and for samples which fall below it) and the increase in overall spread of $\delta v_{\parallel}$ with heliocentric distance indicates that local effects are indeed able to shift these inversions further from the ideal Alfv\'enic case.
Without a second, clearly non-Alfv\'enic, population which is visible in the data, 
the existence of a single generation process for the inversions included in this study is the simplest explanation for these data.
We stress that this argument applies only to the inverted samples which take the form of deviations from a well-defined background field direction, as described in Section \ref{sub:caveats}. Some subset of the samples discarded in constructing this data set could be part of a secondary, entirely non-Alfv\'enic inversion population. However, effective analysis of the inversions in this study has precluded the investigation of this secondary population.
Detailed investigation into the relative prominence of such non-Alfv\'enic inversions will require an alternative approach and is left to future work.

\section{Conclusions}\label{sec:conc}
In this study, we have performed an analysis of the evolving relationship between the deflection angle of the HMF  and the  change in several plasma parameters, particularly the proton velocity vector,  over the entire \textit{Helios} 1 data set, spanning 0.3 to \SI{1}{AU}. We have done so with a particular focus on magnetic inversions, where the deflection magnitude is $>\SI{90}{\degree}$. This type of analysis is possible because of the inclusion of strahl alignment as a means of rectifying samples based on magnetic sector.
A large fraction of inverted HMF samples exhibit an anti-correlated Alfv\'enic relationship between $\delta \mathbf{B}$ and $\delta \mathbf{v}$. 
This result is in general agreement with magnetic reversals and velocity spikes observed previously both near and far from the Sun \citep[][]{Matteini2014,Horbury2018,Kasper2019,Bale2019}.        
However, our results demonstrate that anti-correlated $\delta \mathbf{B}$ and $\delta \mathbf{v}$ for inversions is clear even in solar wind data with minimal discrimination by type; likely a result of observing down to \SI{0.3}{AU} with a large statistical data set, and selecting against samples with a strong non-azimuthal field component. 
The similarity of the inversion properties with those observed inside \SI{0.3}{AU} by PSP suggest the same structures exist over a wide heliocentric distance range. 
However, PSP has observed inversions on time scales smaller than those accessible using the present \textit{Helios} data set.
The prominence of Alfv\'enic switchbacks in the PSP data may be a result of the far greater Alfv\'en speed close to the Sun, which renders the velocity component far more striking than similarly Alfv\'enic inversions far from the Sun \citep[as suggested by e.g.,][]{Matteini2014}.

Examining the properties of inversions at increasing solar distance, $r$, we find that they depart further from the `ideal' Alfv\'enic case of energy balance between magnetic field and plasma further from the Sun. This persists to the point where many inversions are not recognisably Alfv\'enic under certain solar wind conditions, and when approaching \SI{1}{AU}, indicating that Alfv\'enicity is not a universal inversion property.
The strongest departures from ideal Alfv\'enic properties are observed coincident with larger changes in $n$ and $B$, indicating enhanced compressibility.
We find no clear evidence that inversions which are non-Alfv\'enic have different origins to those which are close to ideally Alfv\'enic. We instead favour the simple explanation that all inversions included in this study could be formed by some common process, either locally or at the Sun. 
From the conclusions of \cite{Macneil2020inv}, we favour growing turbulent fluctuations as such a process. Differences in respective $\delta \mathbf{B}$--$\delta \mathbf{v}$ relationships could then arise due to transit effects in different solar wind streams. 
Entirely non-Alfv\'enic inversions, which may exist and be discarded by the cautious analysis procedure in this study, could be produced through alternative mechanisms. This topic is left to future work.

With the return of new  \textit{in situ} plasma composition  data from Solar Orbiter \citep{Muller2013},
analysis of heavy ion composition within magnetic inversions will for the first time be possible over a distance range similar to that covered in this study. 
Composition could be leveraged to examine whether inversions of different dynamic properties (e.g., strongly Alfv\'enic) also have distinct compositional features, which would indicate different source locations or processes \citep[see compositional analysis of large scale inversions at L1 by][]{Owens2020}.

\section*{Acknowledgements}
Work was part-funded by Science and Technology Facilities Council (STFC) grant No.\ ST/R000921/1, and Natural Environment Research Council (NERC) grant No.\ NE/P016928/1. 
RTW is supported by STFC Grant ST/S000240/1.
We acknowledge all members of the  \textit{Helios} data archive team\footnote{http://helios-data.ssl.berkeley.edu/team-members/} who made the  \textit{Helios} data publicly available to the space physics community. We thank David Stansby for producing and making available the  \textit{Helios} proton `corefit' data set\footnote{https://doi.org/10.5281/zenodo.891405}.
This research made use of Astropy,\footnote{http://www.astropy.org} a community-developed core Python package for Astronomy \citep{astropy:2013,astropy2018}. This research made use of HelioPy, a community-developed Python package for space physics \citep{Stansby2019}.
All figures were produced using the Matplotlib plotting library for Python \citep{Hunter2007}. 

\section*{Data Availability Statement}
\textit{Helios} corefit data are accessible through the HelioPy package or available directly at \url{https://helios-data.ssl.berkeley.edu/data/E1_experiment/New_proton_corefit_data_2017/}.
\textit{Helios} electron VDFs are available at \url{https://helios-data.ssl.berkeley.edu/data/E1_experiment/E1_original_data/}. 

\bibliographystyle{mnras}
\bibliography{inversions} 

\appendix

\section{Impact of Errors in Background Field Direction}\label{app:angles}
\begin{figure}
    \centering
    \includegraphics[clip,trim={.3cm 0 .1cm 0},width=0.475\textwidth]{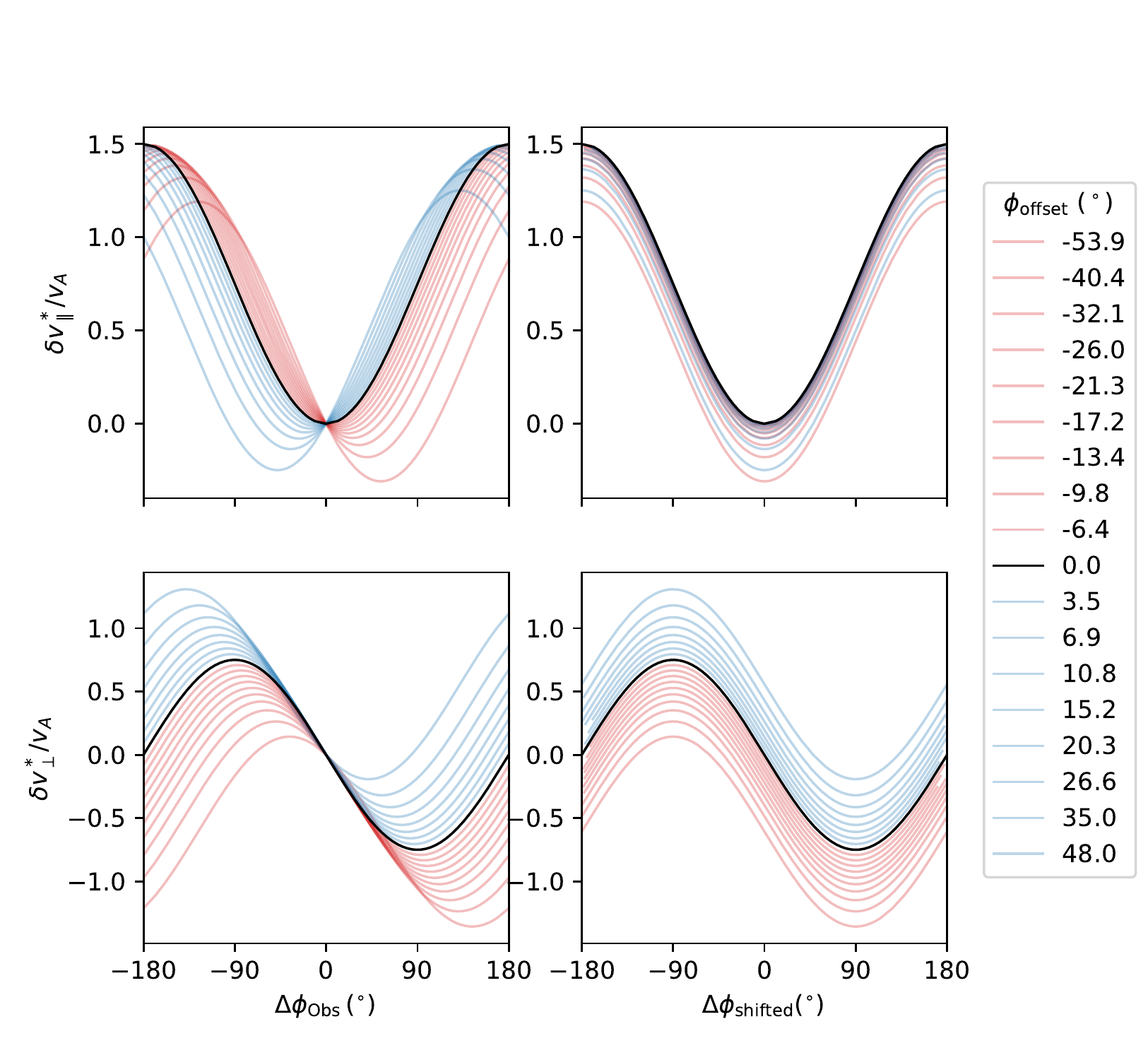}
    \caption{
    Plots of $\delta v^*_{\parallel}/v_A$ (top row) and  $\delta v^*_{\perp}/v_A$ (bottom row) against $\Delta \phi_{\mathrm{obs}}$ (left column) and $\Delta \phi_{\mathrm{shifted}}$, as calculated for different offsets, $\phi_{\mathrm{offset}}$, applied to $\phi_0$ as described in the text. Values of  $\phi_{\mathrm{offset}}$ are based on  the observed distribution $\Delta\phi_P$ for the whole \textit{Helios} 1 data set, ranging from the 5th to 95th percentile values in steps of 5.
    The left column shows $\delta v^*_{\parallel}/v_A$  and  $\delta v^*_{\perp}/v_A$ with the shifts in angle preserved, to reflect the true effects on the observations in e.g., Figure \ref{fig:v_heatmap}. The right column shows the same variables against $\Delta\phi_{\mathrm{shifted}} = \Delta\phi_{\mathrm{obs}}-\phi_{\mathrm{offset}}$, to allow the scales of changes in $\delta v_{\parallel}/v_A$ and $\delta v_{\perp}/v_A$ to be easily compared.
    }
    \label{fig:dvmodel}
\end{figure}

Results of this study are produced under the assumption that the unperturbed background magnetic field angle, $\phi_0$, follows the ideal Parker spiral, $\phi_P$.  Instances of $\phi_0 \neq \phi_P$ will result in an error between the estimated field deflection angle, $\Delta\phi_P$, and the unknown true deflection angle $\Delta\phi$. 
Additionally, parallel and perpendicular proton velocity components  $v_{\parallel}$ and $v_{\perp}$ are calculated relative to the ideal Parker spiral field vector. If the true background magnetic field angle $\phi_0 \neq \phi_P$, then these components will not be aligned correctly with the true background field.
These effects will lead to both an offset in the deflection angle and a shift in the velocity components. To estimate the impact on our results, we construct a simple ideal model of $\delta v_{\parallel (\perp)}$ against  $\Delta \phi_{\mathrm{obs}}$ (observed $\Delta \phi$) for a 2-dimensional Alfv\'enic fluctuation which obeys Equation \ref{eq:dv}. 
Beginning with a unit magnetic field along $\phi_0$,  and an unperturbed velocity vector ($\Delta \phi = \SI{0}{\degree}$), we rotate the field through to $\Delta \phi = \pm\SI{180}{\degree}$ and calculate the resulting velocity perturbation components $\delta v_{\parallel (\perp)}/v_A$, using Equation \ref{eq:dv} with $\sqrt{r_A}=0.75$. The left column of Figure \ref{fig:dvmodel} shows the resulting velocity components against observed $\Delta \phi_{\mathrm{obs}}$ in black.

We next calculate $\delta v^*_{\parallel (\perp)}/v_A$, which represent the observed velocity components for an Alfv\'enic perturbation when the estimated background field direction is different from the true vector about which the perturbation is centred.
To do so for a given offset angle $\phi_{\mathrm{offset}}$, we take $\delta v_{\parallel (\perp)}/v_A$ as calculated above, and project these components onto a new background magnetic field vector, which is rotated by $\phi_{\mathrm{offset}}$, such that $\phi_{\mathrm{obs}} = \phi_0+\phi_{\mathrm{offset}}$. $\phi_{\mathrm{offset}}>0$ means $\phi_{\mathrm{obs}}>\phi_0$, reflecting an ideal Parker spiral angle which is more radial than the true background angle.

The left column of Figure \ref{fig:dvmodel} shows $\delta v^*_{\parallel (\perp)}/v_A$ as calculated for a range of positive and negative $\phi_{\mathrm{offset}}$ shifts. These shifts are chosen using the observed distribution of $\Delta \phi_P$ across all radial distances in the \textit{Helios} 1 data set, by extracting the 5th to 95th percentile values in steps of 5. These values are chosen as a pessimistic estimate of the difference between $\phi_P$ and $\phi_0$, and use percentile values so that regions of the plot where lines are closer together roughly correspond to more probable offsets to $\delta v_{\parallel (\perp)}/v_A$  in the real data.
The right column shows these same $\delta v^*_{\parallel (\perp)}/v_A$ components against $\Delta\phi_{\mathrm{shifted}} = \Delta\phi_{\mathrm{obs}}-\phi_{\mathrm{offset}}$. This accounts for the shift to each $\phi_0$, so that each line is centred around \SI{0}{\degree}. 
The left column of the figure thus represents the effect of these offsets on an observed $\delta v_{\parallel (\perp)}/v_A$-$\Delta\phi_P$ relationship, while the right column isolates the upward and downward shifts of the velocity components, without the associated shift in angle.

Figure \ref{fig:dvmodel} shows that $\delta v^*_{\parallel}$ is shifted down for both signs of $\phi_{\mathrm{offset}}$, while the shift in $\delta v^*_{\perp}$ depends on the sign of $\phi_{\mathrm{offset}}$. The result of this, shown in the top left panel, is that non-zero $\phi_{\mathrm{offset}}$ causes the spread in observed $\delta v_{\parallel}$  to extend more towards values below the Alfv\'enic line, particularly at \SI{0}{\degree} and $\pm\SI{180}{\degree}$. Larger $\phi_{\mathrm{offset}}$ leads to larger offsets in velocity. However, aside from the two most extreme values of $|\phi_{\mathrm{offset}}| \simeq \SI{50}{\degree}$, which we expect to be very uncommon, most offsets will lead to a relatively narrow spread about the Alfv\'enic line in $\delta v^*_{\parallel}$. Conversely, the spread about the Alfv\'enic line for $\delta v^*_{\perp}$ increases with increasing $|\Delta\phi_{\mathrm{obs}}|$, such that the spread in errors in $\delta v^*_{\perp}$ due to an offset field angle will be greatest for larger $\Delta \phi_P$, and smallest near $\Delta \phi_P=\SI{0}{\degree}$.

\section{Additional Inversion Criteria}\label{app:correction}

In Section \ref{sec:intro} we describe an additional criteria which we apply to refine our classification of inverted HMF. In this appendix we shall provide the justification for taking this step. To recap Section \ref{sec:intro}, the first criteria for inverted HMF is that the strahl is oriented in the anti-parallel (parallel) direction for positive (negative) HMF polarity. The second, new, criteria is that the HMF polarity of a given sample be opposite to the modal polarity over a \SI{12}{hr} window centred on that sample. Samples which meet the first criteria but not the second are discarded.

\begin{figure*}
    \centering
    \includegraphics[clip,trim={0 10.05cm 0 0},width=.8\textwidth]{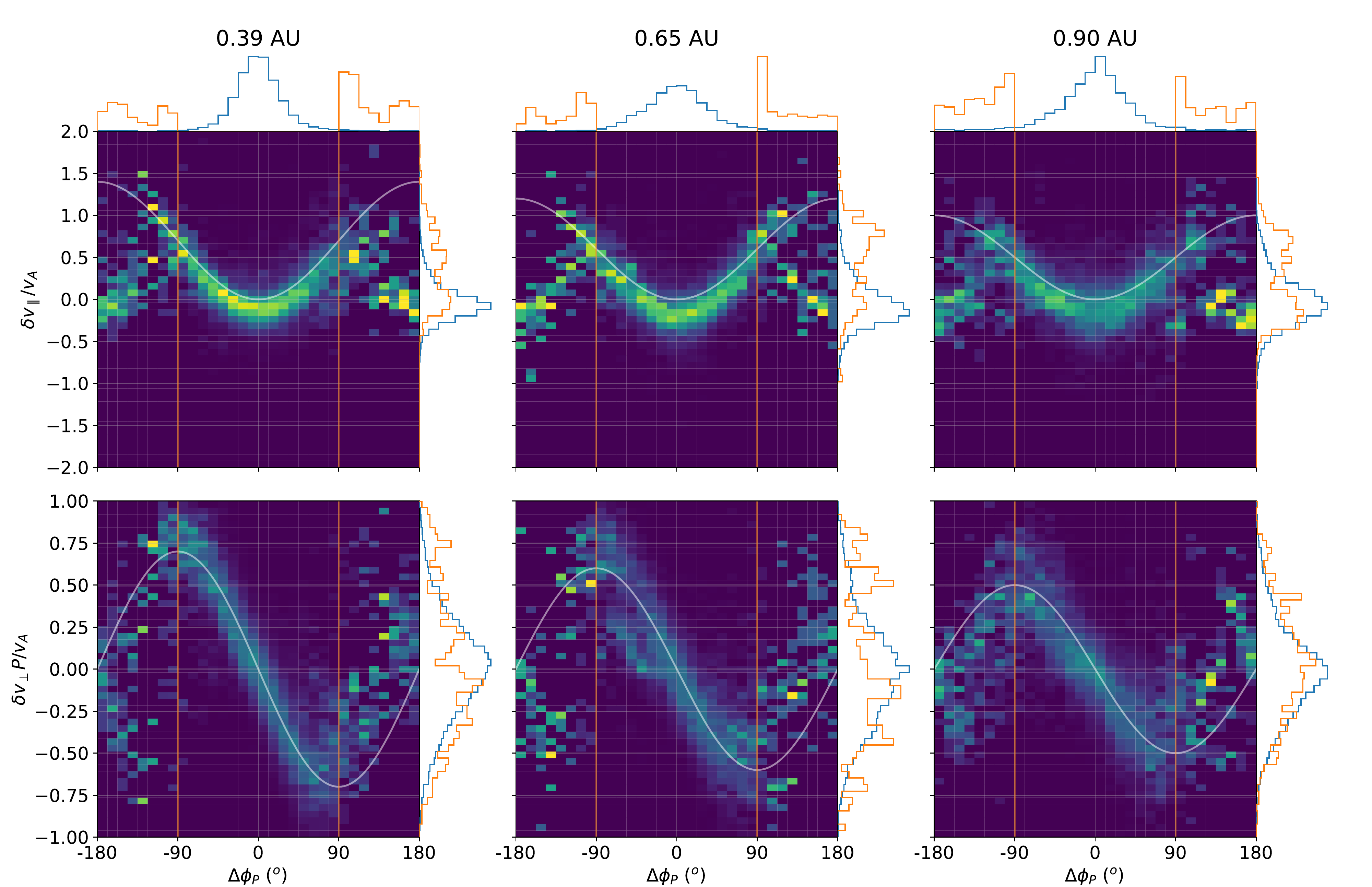}
    \caption{2-dimensional histograms of $ \delta v_{\parallel}/v_A$ and $\delta v_{\perp}/v_A$ against $\Delta \phi_P$ for three radial distance bins in the same format as Figure \ref{fig:vmap_young}. The inverted HMF samples included in this plot ($|\Delta\phi_P|>\SI{90}{\degree}$) are only subject to the first inversion criteria defined by the strahl.  Above the top row of panels, the blue histogram shows the distribution of $\Delta \phi_P$ for the  samples included in the relevant 2-dimensional histograms. The orange histogram shows the distribution  of $\Delta \phi_P$ for the included inverted HMF samples only. The orange and blue histograms are normalised independently. 
    To the right of each panel, normalised histograms show the distribution of that panel's $y$-axis parameter for  uninverted and inverted HMF samples combined (blue line) and for inverted samples only (orange line). }
    \label{fig:vmap_young_example}
\end{figure*}

Figure \ref{fig:vmap_young_example} contains plots identical to Figure \ref{fig:vmap_young}, $\delta v_{\parallel (\perp)}$ for low relative $A_C$ samples, except do not discard  inverted samples which do not meet the second criteria. There is a clear additional population of points, located around $\Delta \phi_P = \pm\SI{180}{\degree}$.  These samples are concentrated around  $\delta v_{\parallel}=0$ at  $\Delta \phi_P = \pm\SI{180}{\degree}$, and increase towards $\Delta \phi_P = \SI{0}{\degree}$. This trend is consistent with an Alfv\'enic relationship, which is found in e.g., Figure \ref{fig:vmap_young}, but shifted by \SI{180}{\degree}. This may be due to incorrect rectification of $\Delta \phi_P$. Mis-rectification would occur when some samples are wrongly identified as inverted HMF. The distributions of $\Delta \phi_P$, shown above the 2-dimensional histograms, also increase towards $\pm\SI{180}{\degree}$, which appears unusual compared to the monotonic decrease which is present for $\Delta \phi_P$ at lower values.

Mis-identification of inverted HMF could occur because the strahl alignment is not being accurately determined, perhaps due to a low overall strahl flux or otherwise anomalous distribution function. Examining electron PADs corresponding to the samples which fall in the additional populations in Figure \ref{fig:vmap_young_example}, we find that a sizeable fraction have a particularly low total strahl flux, which supports this hypothesis. However, removing all samples with a low strahl flux does fully not eliminate the apparently mis-rectified population. Some further inverted samples in this population occur just before or after bidirectional strahl is observed, indicating that the sunward-travelling strahl (which we assume indicates inverted HMF) may be instead due to the appearance of strahl PADs on closed HMF \citep[see][]{Macneil2020inv}.

The second additional criteria for inversions, which we apply for the data used in this study, is designed to mitigate the apparent mis-identification of inversions by applying a separate verification step.
Comparing Figure  \ref{fig:vmap_young_example} to Figure \ref{fig:vmap_young}, we see that the additional population is largely absent, indicating that the second criteria is successful.
The low relative $A_C$ data shown in these figures illustrates the effect of this additional step most clearly, although a similar effect is found when we make equivalent comparisons with e.g., Figures \ref{fig:v_heatmap} and \ref{fig:vmap_old}.

While the mis-rectification of inverted HMF appears to be largely solved by introducing the second criteria, there are further potential consequences of this approach. Primarily, some true HMF inversions likely exist which do not coincide with a polarity which is opposite to the background polarity in a \SI{12}{hr} window. This could be because they occur near the HCS (so the dominant polarity is that of the opposite sector), they are particularly large or occur in bursts (causing the inverted polarity to dominate over a \SI{12}{hr} window), or because the inversion does not coincide with a field deflection visible in the time series, but is clear from the strahl data \citep[such as that shown in Figure 1 of][]{Crooker2004}. This true inverted HMF will be removed by this step in our analysis. The presence of these types of inversion may explain why the removal of only samples with a low strahl flux does not fully remove the non-Alfv\'enic population in Figure \ref{fig:vmap_young_example}.

\bsp	
\label{lastpage}
\end{document}